\documentclass[10pt,twocolumn]{IEEEtran}

\IEEEoverridecommandlockouts

\usepackage[T1]{fontenc}
\usepackage{amssymb,amsmath,epsfig}
\usepackage{array}
\usepackage{booktabs,hyperref}
\usepackage{float,caption,hypcap}
\usepackage[latin5]{inputenc}
\usepackage{graphicx,times,latexsym}
\usepackage{psfrag,cite}
\usepackage{setspace}
\usepackage{algorithmic, algorithm}
\usepackage{times}

\providecommand{\numberTblEq}[1]{\refstepcounter{tblEqCounter}\label{#1}\thetag{\thetblEqCounter}}

\newcommand{\inter}{0.90}
\newcommand{\interDUE}{0.90}
\renewcommand{\baselinestretch}{\inter}
\allowdisplaybreaks

\begin{document}
\newcounter{tblEqCounter}
\title{A Low-Complexity Graph-Based LMMSE Receiver for MIMO ISI Channels with $M$-QAM Modulation}

\author{{\bf Pinar Sen}$^{\S}$, {\bf A. \"{O}zg\"{u}r Y\i lmaz}$^\star$\\
$^{\S}${
University of California San Diego -- La Jolla, CA 92093, USA.} \\
$^\star${Middle East Technical University -- Ankara, Turkey}}

\maketitle

\begin{abstract}
In this paper, we propose a low complexity graph-based linear minimum mean square error (LMMSE) equalizer in order to remove inter-symbol and inter-stream interference in multiple input multiple output (MIMO) communication. The proposed state space representation inflicted on the graph provides linearly increasing computational complexity with block length. Also, owing to the Gaussian assumption used in the presented cycle-free factor graph, the complexity of the suggested equalizer structure is not affected by the size of the signalling space. In addition, we introduce an efficient way of computing extrinsic bit log-likelihood ratio (LLR) values for LMMSE estimation compatible with higher order alphabets which is shown to perform better than the other methods in the literature. Overall, we provide an efficient receiver structure reaching high data rates in frequency selective MIMO systems whose performance is shown to be very close to a genie-aided matched filter bound through extensive simulations. 
\end{abstract}


\IEEEpeerreviewmaketitle

\IEEEpubidadjcol

\section{INTRODUCTION}
\label{sec:intro}

MIMO systems have attracted much attention in recent years since they potentially provide high spectral efficiency in wireless communication applications. Yet, they require complicated receiver structures so as to handle the distortion caused by the wireless channel characteristics such as the intersymbol interference (ISI) resulting from the frequency selectivity of the channel between each transmit and receive antenna pair.

In recent studies, low complexity equalizer structures are proposed to mitigate those distorting effects in MIMO ISI channels. Although, frequency domain (FD) approaches hold an important place in the literature~\cite{Gokhan2013,Murat2008,Falconer2002,Yuan2008}, due to the problems related to FD methods, low complexity time domain approaches have drawn interest from the perspective of the lately studied factor graph theory~\cite{Loeliger2002,Loeliger2004,Loeliger2006,Loeliger2007,Sascha2005,Singer2007}. Belief propagation and sum product algorithms on factor graphs were proposed for both single input single output (SISO) and MIMO systems~\cite{Colavolpe2005,Duman2007}, but they have $O(M^{\tilde{P}})$ complexity per symbol where $M$ is the constellation size and $\tilde{P}$ is the total number of \textit{non-zero} interferers. 

The Gaussian assumption utilized in the equalizer structures which provides constant complexity with increasing alphabet size has become popular lately. As an example, Kalman filtering was proposed for coded frequency selective MIMO systems in~\cite{Duman20072}. However, it has $O(P^3)$ complexity per symbol where $P$ is the number of interferers, and more importantly lacks the improvement that backward recursion provides. On the other hand, the Gaussian message passing (GMP) rules including Kalman filtering (forward recursion) and Kalman smoothing (backward recursion) operations are derived~\cite{Sascha2005,Loeliger2006} and used in the implementation of LMMSE equalization on factor graphs. This approach has the advantage of complexity linearly increasing with block length $N$ as compared to conventional block LMMSE filter's $O(N^3)$ complexity~\cite{Tuchler2002b}. Although the factor graph structures with cycles using the GMP rules were proposed for SISO and MIMO ISI channels respectively in~\cite{Colavolpe2011,Springer2012}, our main focus is the cycle free ones due to exact equivalence to LMMSE filtering avoiding any iterations. There are two different cycle free factor graph structure presented in the literature for SISO systems~\cite{Singer2007,Guo2008}. The generalization of~\cite{Singer2007} to MIMO ISI channels was proposed in~\cite{Springer2011} which still has $O(P^3)$ complexity per symbol. Also, the mentioned studies including the GMP rules do not have any performance results for modulation types other than BPSK signaling due to the lack of LLR exchange algorithm. 

In this study, however, we reduce the complexity to $O(P^2)$ per symbol with the help of a factor graph structure which takes its roots from~\cite{Guo2008}. Moreover, using Gaussian approximation of GMP rules keeps the complexity of the graph algorithm constant with the increasing constellation size. In addition, the presented approach here brings the ease of involving existing \textit{a priori} information of the transmitted symbols, hence perfectly matched with the turbo concept for coded systems. It is also well suited to fast fading environments since the channel taps (possibly time-varying) are directly included in the graph. Therefore, the proposed structure is a very advantageous way of implementing LMMSE filtering for equalization of MIMO ISI channels.

Another important contribution of this study is the proposed LLR exhange algorithm for $M$-QAM signaling. LMMSE equalizers involved in turbo decoders need a method for transition to binary domain, i.e., bit LLR domain. In the literature, there were effective approaches to obtain bit LLRs from the LMMSE equalizer outputs, such as the Wang-Poor (WP) approach~\cite{WangPoor1999,Tuchler2002} and the Joint Gaussian (JG) approach~\cite{Ping2003}. However, applying the WP or JG approaches directly is computationally intensive for factor graphs. Although a simplified expression for extrinsic LLR computation was proposed in~\cite{Guo2008} for BPSK signaling only, there is no such a work for higher order constellations in the literature within the knowledge of the authors except the heuristic methods in~\cite{Guo2011,Pinar2014}. To fill up this gap, we derive a transformation from the graph outputs to the bit LLRs based on the WP approach for higher order modulation alphabets. Owing to this key connection, extrinsic bit LLR values can be obtained in accordance with the graph solution without major complexity increase.

In summary, two main contributions of this study are: i) a state space graph for time domain LMMSE equalization of MIMO ISI channels with reduced complexity as compared to the techniques in the literature,
ii) a computationally simple method to obtain extrinsic bit LLRs from LMMSE equalizer outputs for $M$-QAM signaling in SISO and MIMO systems.

Overall, the performance of the proposed extrinsic bit LLR producing algorithm is shown to be better as compared to the heuristic methods in the literature for $M$-QAM signaling. Also, the performance of the extended LMMSE equalizer using this LLR producing algorithm is shown to be very close to a genie-aided matched filter bound~\cite{Barry2004} through extensive simulations which makes it an efficient receiver that can reach high data rates in frequency selective MIMO systems.

The paper is organized as follows. Section~\ref{sec:system_model} gives the system model. Section~\ref{sec:Equalizer_MIMO} presents our proposed factor graph-based LMMSE equalizer design. In Section~\ref{sec:llr_exchange}, the proposed LLR exchange algorithm is analyzed in details. We discuss the computational complexity of the suggested receiver in Section~\ref{sec:complexity}. Section~\ref{sec:Simulation} presents the bit error rate (BER) performance results of the proposed receiver structure. Lastly, Section~\ref{sec:conclsn} concludes the paper.

\section{SYSTEM MODEL}
\label{sec:system_model}
The notations used in the paper are organized as follows. Lower case letters (e.g., $x$) denote scalars, lower case bold letters (e.g., $\mathbf{x}$) denote vectors, upper case bold letters (e.g., $\mathbf{X}$) denote matrices. For a given random variable $x$; $m_x$, $v_x$, $w_x$ and $w_x m_x$ denote its mean, variance, weight and weighted mean values respectively where $w_x \triangleq v_x^{-1}$. For a given vector random variable $\mathbf{x}$; $\mathbf{m_x},\mathbf{V_x},\mathbf{W_x}$ and $\mathbf{W_x}\mathbf{m_x}$ denote 
its mean vector, covariance matrix, weight matrix and weighted mean vector respectively where $\mathbf{W_x} \triangleq \mathbf{V_x}^{-1}$. The indicators $()^T$, $()^H$, and $E\{\}$ denote transpose, Hermitian transpose and expectation respectively and $\mathbf{I}$ denotes the identity matrix of proper size. $diag(\mathbf{A})$ is defined as the diagonal elements of $\mathbf{A}$ and $diagMat(\mathbf{a})$ is defined as the diagonal matrix with $diag(diagMat(\mathbf{a}))=\mathbf{a}$. $blkdiag([\mathbf{A}_1,\mathbf{A}_2,\ldots,\mathbf{A}_n])$ denotes the block diagonal matrix where $i^{th}$ main diagonal matrix is $\mathbf{A}_i$. Lastly, $Toeplitz(\mathbf{A})$, for $\mathbf{A}=[\mathbf{A}_1,\mathbf{A}_2,\ldots,\mathbf{A}_n]$ is defined as 
\begin{footnotesize}
\begin{align*}
Toeplitz(\mathbf{A}) = \left[\begin{array}{ccccccc}
\mathbf{A}_1 & \mathbf{0} & \mathbf{0}& \mathbf{0} & \mathbf{0} \\
\vdots & \ddots & \ddots & \vdots & \vdots \\
\mathbf{A}_n & \mathbf{A}_{n-1}  & \ldots &  \mathbf{A}_{1} & \mathbf{0} \\ \vdots & \ddots & \ddots & \vdots & \vdots \\
\mathbf{0} & \mathbf{0} &  \ldots &  \mathbf{0} & \mathbf{A}_{n} 
\end{array} \right].
\end{align*}
\end{footnotesize}
We consider a MIMO single-carrier communication system which suffers from the ISI effect due to the wireless nature of the channel. The block diagram of the discussed transmitter and receiver structures are depicted in Fig.~\ref{fig:system_model_MIMO}. At the transmitter side, after the coded information bits are interleaved and modulated according to an $M$-QAM alphabet $S$, modulated symbols are spread to $N_t$ transmit antennas and sent over the ISI channel which occurs between each transmit and receive antenna. At the receiver, a turbo structure including the proposed graph based LMMSE equalizer and \textit{a posteriori} probability (APP) decoder is operated by use of observations from $N_r$ receive antennas. One turbo iteration is defined as one cycle of consecutive operations of equalizer and APP decoder.

We can model the given discrete-time system at time $k$ as
\begin{align}
\label{eqn:observation}
\mathbf{y}_k &= \sum_{i=0}^{L-1} \mathbf{H}_i \mathbf{x}_{k-i} + \mathbf{n}_k \,;\quad k = 1,2,\ldots, N+L-1 , \\ \nonumber
\mathbf{H}_i &= \left[ \begin{array}{cccc}
h_{11}(i) & h_{12}(i) & \ldots & h_{1 N_t}(i)\\
h_{21}(i) & h_{22}(i) & \ldots & h_{2 N_t}(i)\\
\vdots & \vdots & \vdots & \vdots\\
h_{N_r 1}(i) & h_{N_r 2}(i) & \ldots & h_{N_r N_t}(i)\\
\end{array} \right]_{N_r\times N_t} ; 
\end{align}
and $L$ is the number of channel taps; $N$ is the transmission block length; $\mathbf{H}_i$ is the $N_r \times N_t$ channel matrix at time $i$; $\mathbf{x}_{k}$ is the transmitted symbol vector of size $N_t$ at time $k$; $\mathbf{y}_k$ is the observation vector of size $N_r$ at time $k$ and $\mathbf{n}_k$ represents additive white circularly symmetric complex Gaussian noise vector with zero mean and covariance $N_0\mathbf{I}_{N_r}$ at time $k$, i.e., $\mathbf{n}_k \sim CN(0,N_0 \mathbf{I}_{N_r})$. The input symbol sequence is assumed to have independent, identically distributed (i.i.d.) random variables and the transmitted symbol vector $\mathbf{x}_{k}$ at time $k$ is $\mathbf{x}_k = [x_{k,1} \; x_{k,2} \; \ldots \; x_{k,N_t}]^T$ for $k=1,2, \ldots N$ where $x_{k,l}$ is the symbol transmitted at the $l^{th}$ transmit antenna at time $k$ and its average energy is defined as $E_s$, i.e., $E\{\vert{x_{k,l}}\vert^2\} \triangleq E_s$. For notational convenience, we define $J \! \triangleq \! L-1$ which denotes the memory of the channel.
\begin{figure}[t]
   \centering
   \includegraphics[width=0.45\textwidth]{figures/system_model_MIMO_ISI}
   \caption{System Model}
   \label{fig:system_model_MIMO}
\end{figure}

Considering the multiplexing operation in Fig.~\ref{fig:system_model_MIMO}, matrix representation of~(\ref{eqn:observation}) can be written as
$\mathbf{y} = \mathbf{H} \mathbf{x} + \mathbf{n}$, where $\mathbf{y} = \left[\mathbf{y}_1^T \; \mathbf{y}_2^T \; \ldots \; \mathbf{y}_{N+J}^T \right]^T$,   $\mathbf{x} = \left[ 
\mathbf{x}_1^T \; \mathbf{x}_2^T \; \ldots \; \mathbf{x}_N^T \right]^T$, $\mathbf{n} = \left[ \mathbf{n}_1^T \; \mathbf{n}_2^T  \; \ldots \; \mathbf{n}_{N+J}^T \right]^T$, and $\mathbf{H} = Toeplitz([\mathbf{H}_0 \: \mathbf{H}_1 \: \ldots \:\mathbf{H}_J])$.

For the described system model, the details of the proposed equalizer structure are given in the subsequent section.

\section{GRAPH BASED LMMSE EQUALIZER FOR MIMO ISI CHANNEL}
\label{sec:Equalizer_MIMO}

\begin{figure}[t]
   \centering
   \includegraphics[width=0.5\textwidth]{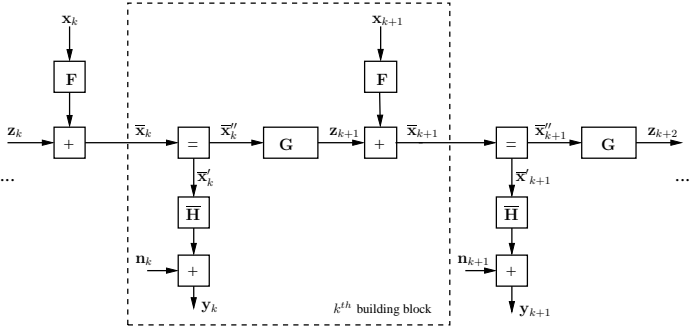}
   \caption{Factor Graph of MIMO ISI Channel}
   \label{fig:LMMSE_FG_MIMO}
\end{figure}

In this section, we elucidate the proposed graph structure together with the message passing algorithm. Construction of the proposed graph takes its roots from the state space representation of LMMSE equalization for SISO systems in~\cite{Loeliger2004,Guo2008}. In fact, the authors of~\cite{Singer2007} and~\cite{Springer2011} construct another factor graph for LMMSE equalization in SISO and MIMO systems respectively based on factorizations. However, the state space representation has couple of advantages over the latter approach. First, it is computationally more efficient. As to be shown, one can compose adjacent blocks with the help of the matrix inversion lemma and reduce the computational complexity to $O(P^2)$ per symbol where $P$ is the number of interferers; whereas, this type of composition is not natural for the graph structure in~\cite{Singer2007,Springer2011}, that results in complexity of $O(P^3)$ per symbol. Moreover, the flow of the messages are easy to follow on the proposed graph, where operations on each building block, shown in Fig.~\ref{fig:LMMSE_FG_MIMO}, are identical to each other. On the other hand, messages in~\cite{Singer2007,Springer2011} piece-wisely defined on three different regions, which further complicates implementation. Hence, in this study, a graph structure is constructed using state space representation. The GMP rules generated for the graph implementation of LMMSE estimation in~\cite{Loeliger2002,Loeliger2004,Loeliger2006,Loeliger2007} are operated on the constructed graph in which all state variables are assumed to have Gaussian distribution. Therefore, each state variable is represented by a mean and variance value on the graph which makes it a suitable receiver for higher order constellations. Before going into the details of message passing rules, we begin with the state space representation of the system presented in Section~\ref{sec:system_model} to construct the graph structure. For the system described in Fig.~\ref{fig:system_model_MIMO}, the observation vector at time $k$ given in (\ref{eqn:observation}) can be rewritten as
\begin{align}
\label{eqn:observation_for_LMMSE}
\mathbf{y}_k &= \overline{\mathbf{H}} \; \overline{\mathbf{x}}_k + \mathbf{n}_k \; k=1,2, \ldots ,N+J, \; \text{where} \\ 
\overline{\mathbf{H}} &= [\mathbf{H}_{J} \: \mathbf{H}_{J-1}  \ldots  \mathbf{H}_0], \;
\label{eqn:chan_matrix_LMMSE_x}
\overline{\mathbf{x}}_k  \!= \!\! \left[ \mathbf{x}_{k-J}^T \; \mathbf{x}_{k-J+1}^T \; \ldots \; \mathbf{x}_{k}^T  \right]^T\!\!\!.
\end{align}   
We use (\ref{eqn:observation_for_LMMSE})-(\ref{eqn:chan_matrix_LMMSE_x}) to construct the state space graph representation of the MIMO ISI channels similar to~\cite{Guo2008} which discusses the SISO ISI channel case. For transitions to the next time instant, $k+1$, we define
\begin{align*}
\mathbf{G} =& \left[ \begin{array}{cc}
\mathbf{0}_{N_t J \times N_t} & \mathbf{I}_{N_t J} \\
\mathbf{0}_{N_t \times N_t} & \mathbf{0}_{N_t \times N_t J} \\
\end{array} \right],\; \mathbf{F} =& \left[ \begin{array}{cc}
\mathbf{0}_{N_t J \times N_t} \\
\mathbf{I}_{N_t}\\
\end{array} \right]
\end{align*}
where $\mathbf{0}$ denotes the all zero matrix of the specified size and $\mathbf{I}_j$ denotes the identity matrix of size $j$. It can be seen that
\begin{align}
\overline{\mathbf{x}}_{k+1} &= \mathbf{F}\;\mathbf{x}_{k+1} + \mathbf{z}_{k+1}, \quad \text{where} \\
\mathbf{z}_{k+1} &= \mathbf{G}\; \overline{\mathbf{x}}_{k} = \left[ 
\mathbf{x}_{k-J+1}^T\:
\mathbf{x}_{k-J+2}^T\:
\ldots\:
\mathbf{x}_{k}^T \:
\mathbf{0}_{1 \times N_t} \right]^T.
\label{eqn:FG_state_defn_z}
\end{align}

The factor graph representation corresponding to (\ref{eqn:observation_for_LMMSE})-(\ref{eqn:FG_state_defn_z}) can be seen in Fig.~\ref{fig:LMMSE_FG_MIMO}. LMMSE equalization is performed on this graph with the help of the GMP rules which are first proposed in~\cite{Loeliger2002} and later discussed in~\cite{Loeliger2004,Loeliger2006,Loeliger2007,Guo2008}. Some of the state variable vectors on the graph are named as shown in Fig.~\ref{fig:LMMSE_FG_MIMO} (such as ${\overline{\mathbf{x}}_{k}},\overline{\mathbf{x}}_{k}^{'}, \overline{\mathbf{x}}_{k}^{''},{\mathbf{z}}_{k}$ and etc.) to help explain the algorithm clearly. Each state variable vector on the factor graph is assumed to have Gaussian distribution and represented by a mean vector $(\mathbf{m}_{\overline{\mathbf{x}}_{k}})$ and a covariance matrix $(\mathbf{V}_{\overline{\mathbf{x}}_{k}})$. \textit{A posteriori} mean $(\mathbf{m}_{\overline{\mathbf{x}}_{k}}^{post})$ and covariance $(\mathbf{V}_{\overline{\mathbf{x}}_{k}}^{post})$ of the state variables are calculated through the GMP rules which are applied in forward and backward recursions by use of the observations $(\mathbf{y})$ and the \textit{a priori} information $(\mathbf{m}_{\mathbf{x}_k}^{\downarrow}$,$\mathbf{V}_{\mathbf{x}_k}^{\downarrow})$ coming from the APP decoder. In Table~\ref{tab:table1} (in Appendices), some of the GMP rules for basic blocks~\cite{Loeliger2006,Guo2008} are provided for self-containment. Those rules could be directly applied to the building blocks of the graph in Fig.~\ref{fig:LMMSE_FG_MIMO}. 

However, the direct application results in quite a few $N_t L$-size matrix inversions each of which costs $O(N_t^3 L^3)$. Hence, we also list the GMP rules for composite blocks some of which are derived in~\cite{Loeliger2004,Loeliger2006} and some of which are obtained by matrix inversion lemma~\cite{Johnson1990} in the last two columns of Table~\ref{tab:table1} to reduce the computational complexity. A brief description of the forward and backward recursion algorithms is provided below for the $k^{th}$ building block. The arrows are used so as to show the direction of the messages as a similar notation to~\cite{Loeliger2004,Loeliger2006,Loeliger2007,Guo2008}. 

\noindent\textbf{\textsl{Forward Recursion:}} We aim to reach the information related to the state $\overline{\mathbf{x}}_{k+1}$ by use of the known values of the state $\overline{\mathbf{x}}_{k}$ obtained by the previous building block and the operations given below. Following the direction from left to right on the $k^{th}$ building block of the graph in Fig.~\ref{fig:LMMSE_FG_MIMO}, we compute $\overrightarrow{\mathbf{m}}_{\overline{\mathbf{x}}^{''}_k}$ and $\overrightarrow{\mathbf{V}}_{\overline{\mathbf{x}}^{''}_k}$ by using $\overrightarrow{\mathbf{m}}_{\overline{\mathbf{x}}_k}$, $\overrightarrow{\mathbf{V}}_{\overline{\mathbf{x}}_k}$ coming from the previous building block and the observation vector $\mathbf{y}_k$ through \eqref{composite_forw_v}-\eqref{composite_forw_m}. To get $\overrightarrow{\mathbf{m}}_{\mathbf{z}_{k+1}}$ and $\overrightarrow{\mathbf{V}}_{\mathbf{z}_{k+1}}$, we use \eqref{matf_v}. With the obtained $\overrightarrow{\mathbf{m}}_{\mathbf{z}_{k+1}}$, $\overrightarrow{\mathbf{V}}_{\mathbf{z}_{k+1}}$ values and the \textit{a priori} information provided by the APP decoder $(\mathbf{m}^\downarrow_{\mathbf{x}_{k+1}}$, $\mathbf{V}^\downarrow_{\mathbf{x}_{k+1}})$, the mean and variance values of the state vector $\overline{\mathbf{x}}_{k+1}$ are computed by \eqref{sum_v},\eqref{matf_v} and used in the next building block as input. By repeating this process for all the building blocks in a serial order, forward recursion is completed.

\noindent\textbf{\textsl{Backward Recursion:}} In each building block, the purpose is to obtain the weight matrix and the weighted mean vector of the state $\overline{\mathbf{x}}_{k}$ from the known information related to the state $\overline{\mathbf{x}}_{k+1}$ provided by the previous building block. Following the direction from right to left, first we compute $\overleftarrow{\mathbf{W}}_{\mathbf{z}_{k+1}}$, $\overleftarrow{\mathbf{W}}_{\mathbf{z}_{k+1}} \overleftarrow{\mathbf{m}}_{\mathbf{z}_{k+1}}$ through \eqref{composite_back_v}-\eqref{composite_back_m} with the help of the \textit{a priori} information coming from the APP decoder $(\mathbf{m}^\downarrow_{\mathbf{x}_{k+1}}$, $\mathbf{V}^\downarrow_{\mathbf{x}_{k+1}})$ and the obtained information of the state $\overline{\mathbf{x}}_{k+1}$ $(\overleftarrow{\mathbf{W}}_{\overline{\mathbf{x}}_{k+1}}$, $\overleftarrow{\mathbf{W}}_{\overline{\mathbf{x}}_{k+1}} \overleftarrow{\mathbf{m}}_{\overline{\mathbf{x}}_{k+1}})$ by the previous building block. Then, after $\overleftarrow{\mathbf{W}}_{\overline{\mathbf{x}}^{''}_k}$ and $\overleftarrow{\mathbf{W}}_{\overline{\mathbf{x}}^{''}_k} \overleftarrow{\mathbf{m}}_{\overline{\mathbf{x}}^{''}_k}$ are computed by \eqref{matb_v}, they are utilized in \eqref{equal_w}-\eqref{equal_m},\eqref{matb_v} together with the observation vector $\mathbf{y}_{k}$ so as to reach $\overleftarrow{\mathbf{W}}_{\overline{\mathbf{x}}_{k}}$, $\overleftarrow{\mathbf{W}}_{\overline{\mathbf{x}}_{k}} \overleftarrow{\mathbf{m}}_{\overline{\mathbf{x}}_{k}}$. These operations are applied to each building block serially in a similar way to forward recursion except message passing direction.

When forward and backward recursion is completed, the output mean vector and covariance matrix of each state vector $\overline{\mathbf{x}}_{k}$ are calculated with the help of the obtained $(\overrightarrow{\mathbf{V}}_{\overline{\mathbf{x}}_{k}},\overrightarrow{\mathbf{m}}_{\overline{\mathbf{x}}_{k}})\text{ and }(\overleftarrow{\mathbf{W}}_{\overline{\mathbf{x}}_{k}},\overleftarrow{\mathbf{W}}_{\overline{\mathbf{x}}_{k}}\overleftarrow{\mathbf{m}}_{\overline{\mathbf{x}}_{k}})$
as in~\cite{Loeliger2002},\cite{Guo2008}:
\begin{align}
\label{eqn:overall_v}
\mathbf{V}_{\overline{\mathbf{x}}_k}^{post} =& (\overrightarrow{\mathbf{V}}^{-1}_{\overline{\mathbf{x}}_{k}} + \overleftarrow{\mathbf{W}}_{\overline{\mathbf{x}}_{k}})^{-1},\\
\label{eqn:overall_m}
\mathbf{m}_{\overline{\mathbf{x}}_k}^{post} =& \mathbf{V}_{\overline{\mathbf{x}}_k}^{post} (\overrightarrow{\mathbf{V}}^{-1}_{\overline{\mathbf{x}}_{k}} \overrightarrow{\mathbf{m}}_{\overline{\mathbf{x}}_{k}} + \overleftarrow{\mathbf{W}}_{\overline{\mathbf{x}}_{k}}\overleftarrow{\mathbf{m}}_{\overline{\mathbf{x}}_{k}})^{-1}.
\end{align}

\noindent\textbf{{Proposition 1:}} $\mathbf{V}_{\overline{\mathbf{x}}_k}^{post}$, $\mathbf{m}_{\overline{\mathbf{x}}_k}^{post}$ given in (\ref{eqn:overall_v}),(\ref{eqn:overall_m}) in this paper are equal to $\mathbf{\Sigma}_{11}^{(k)^{-1}}$, $\mathbf{\mu}_{11}^{(k)}$ in (27),(28) in~\cite{Springer2011}, respectively. In other words, both our graph and the one in~\cite{Springer2011} implement LMMSE equalization, although they have different internal operations; i.e., internal messages do not trivially coincide with each other. The proof of Proposition 1 is given in Appendices.

The diagonal elements of $\mathbf{V}_{\overline{\mathbf{x}}_k}^{post}$ give the \textit{a posteriori} variance values of the symbols sent from all transmit antennas between the time instants $k-J$ and $k$ as given by
\begin{align*}
&\quad\quad\mathbf{V}_{\overline{\mathbf{x}}_{k}}^{post} = blkdiag\left(\left[ 
\mathbf{V}_{\mathbf{x}_{k-J}}^{post} \; \mathbf{V}_{\mathbf{x}_{k-J+1}}^{post} \; \ldots \; \mathbf{V}_{\mathbf{x}_{k}}^{post}  \right]\right),
\end{align*}
where we have
\begin{align*}
&diag(\mathbf{V}_{\overline{\mathbf{x}}_k}^{post}) = [diag( \mathbf{V}_{\mathbf{x}_{k-J}}^{post}) \; diag( \mathbf{V}_{\mathbf{x}_{k-J+1}}^{post}) \ldots diag( \mathbf{V}_{\mathbf{x}_{k}}^{post})], \\
&diag( \mathbf{V}_{\mathbf{x}_{k}}^{post}) = [v_{x_{k,1}}^{post} \; v_{x_{k,2}}^{post} \: \ldots \: v_{x_{k,N_t}}^{post}].
\end{align*}
In a similar way, the elements of $\mathbf{m}_{\overline{\mathbf{x}}_k}^{post}$ includes the \textit{a posteriori} mean values of the state vector $\overline{\mathbf{x}}_{k}$ as below:
\begin{small}
\begin{align*}
\mathbf{m}_{\overline{\mathbf{x}}_k}^{post} &= \left[ \left(\mathbf{m}_{\mathbf{x}_{k-J}}^{post}\right)^T \; \left(\mathbf{m}_{\mathbf{x}_{k-J+1}}^{post}\right)^T \; \ldots \; \left(\mathbf{m}_{\mathbf{x}_{k}}^{post} \right)^T \right]^T \; \text{, where} \\ 
\mathbf{m}_{\mathbf{x}_{k}}^{post} &= \left[ m_{x_{k,1}}^{post} \; m_{x_{k,2}}^{post} \; \ldots \; m_{x_{k,N_t}}^{post} \right]^T.
\end{align*}
\end{small}
Since the elements of the state vector $\overline{\mathbf{x}}_{k}$ is shifted by $N_t$ symbols through the way to $\overline{\mathbf{x}}_{k+1}$, this shift is also seen at the output mean vectors and variance matrices as below:
\begin{small}
\begin{align}
\label{eqn:mv_shift}
\mathbf{V}_{\overline{\mathbf{x}}_{k+1}}^{post} &= blkdiag \left( \left[  \mathbf{V}_{\mathbf{x}_{k-J+1}}^{post} \; \mathbf{V}_{\mathbf{x}_{k-J+2}}^{post} \; \ldots \; \mathbf{V}_{\mathbf{x}_{k+1}}^{post} \right]\right), \\
\mathbf{m}_{\overline{\mathbf{x}}_{k+1}}^{post} &= \left[ \left( \mathbf{m}_{\mathbf{x}_{k-J+1}}^{post}\right)^T \: \left( \mathbf{m}_{\mathbf{x}_{k-J+2}}^{post} \right)^T \: \ldots \: \left( \mathbf{m}_{\mathbf{x}_{k+1}}^{post} \right)^T 
\right]^T
\end{align}
\end{small}
It should be noted that the symbols sent from different transmit antennas are assumed to be independent. So, the \textit{a priori} information related to $\mathbf{x}_k$ is involved in the factor graph as 
\begin{align*}
\mathbf{m}^\downarrow_{\mathbf{x}_{k}} & \triangleq \mathbf{m}^{prio}_{\mathbf{x}_{k}} = \left[m^{prio}_{x_{k,1}} \; m^{prio}_{x_{k,2}} \; \ldots \; m^{prio}_{x_{k,N_t}} \right]^T  , \\ \nonumber 
\mathbf{V}^\downarrow_{\mathbf{x}_{k}} & \triangleq \mathbf{V}^{prio}_{\mathbf{x}_{k}} = diagMat \left(\left[ 
v^{prio}_{x_{k,1}} \;  v^{prio}_{x_{k,2}} \; \ldots \; v^{prio}_{x_{k,N_t}} \right] \right)
\end{align*}
where $m^{prio}_{x_{k,l}}$ and $v^{prio}_{x_{k,l}}$ are the mean and variance values computed under the Gaussian assumption by using the LLR values obtained by APP decoder.

An algorithm is needed to convert the output of the LMMSE equalizer, which is in the form of mean and variance values at this point, to the extrinsic bit LLRs. In the next section, we propose an algorithm consistent with the factor graph to maintain the low complexity for higher order alphabets.

\section{LLR EXCHANGE ALGORITHM COMPATIBLE WITH THE GRAPH APPROACH}
\label{sec:llr_exchange}

LMMSE equalizer used in turbo decoders needs an algorithm to transit between binary, i.e., bit LLR domain, and Gaussian domain. Transition from binary to Gaussian domain is rather trivial and can be reached in equations (2.28-2.29) in~\cite{Pinar2014Tez}. On the other hand, there are mathematical models for the extrinsic bit LLR computation of the LMMSE equalizer in the literature such as the WP~\cite{WangPoor1999,Tuchler2002} and the JG approaches~\cite{Ping2003} which are not suitable for the graph based LMMSE equalization due to their high computational complexity caused by matrix inversions of size $N_t N$. In~\cite{Guo2008}, considering the graph outputs, the mathematical expression of the extrinsic bit LLRs with respect to the JG approach was simplified for BPSK signaling. Also, the authors of~\cite{Guo2008} shows the equivalence between the JG and WP approaches for BPSK signaling. However, there is no mathematically justified reduced complexity LLR exchange algorithm for higher constellation sizes in the literature to the best of our knowledge. Despite the fact~\cite{Guo2011} proposed an intuitive method for $M$-QAM signaling without any simulation results, we have observed that equation $(8)$ in~\cite{Guo2011} causes both diversity and SNR losses as shown in Section~\ref{sec:simple_WP}. This performance loss is because equation $(8)$ in~\cite{Guo2011} depends on the assumption that $p(x_{k,j} \vert \mathbf{y})$ has a Gaussian distribution. In addition, we have also proposed a heuristic algorithm in which both the intrinsic and the a priori LLRs are computed under the Gaussian assumption presented in~\cite{Pinar2014}. Although it has much better performance than the one in~\cite{Guo2011} for M-QAM signaling, there exists no scientifically proved basis for the idea behind our heuristic method. Another method, used in~\cite{Springer2011}, computes the extrinsic information (in terms of mean and variance) in Gaussian domain and obtains the extrinsic LLRs using this information. We call it as LMMSE-EG in the simulation results. The reason why this method fails is that it depends on the assumption where both $p(x_{k,j} \vert \mathbf{y})$ and $p(x_{k,j})$ has Gaussian distribution. On the other hand, WP algorithm, which is to be analyzed in details, works under the assumption that the filtered output has a conditional Gaussian distribution given the actual symbol transmitted~\cite{WangPoor1999}. So, the residual interference plus noise terms is assumed to have a Gaussian distribution, which turns out to be a better and consistent assumption shown by extensive simulations in this paper. Consequently, we will base our proposal on the WP approach which is observed to perform better for $M$-QAM signaling as compared to the others.

In the subsequent section, we provide the mathematical relation between the graph based LMMSE equalizer outputs (\textit{a posteriori} mean and variance values) and the bit LLRs for higher order modulation alphabets. Hence, owing to this key connection, extrinsic bit LLR values from LMMSE estimation can be obtained easily in accordance with the graph solution without any major complexity increase. 

\subsection{Simplified WP Approach for Graph Based LMMSE}
\label{sec:simple_WP}

WP approach is a famous extrinsic bit LLR computation method for LMMSE estimation which was first presented in~\cite{WangPoor1999} and later proposed to be used in iterative decoder structures in~\cite{Tuchler2002} for SISO systems. Since the input symbols are independent and identically distributed, the multiple number of transmit and receive antennas results in just an enlargement in signalling space and does not pose a problem to utilize WP approach for our MIMO system. 

For clear understanding, we can rewrite the observation vector as $\mathbf{y} = \sum_{k=1}^{N} \sum_{j=1}^{N_t} \mathbf{h}_{k,j} \: x_{k,j} + \mathbf{n}$ where $\mathbf{h}_{k,j}$ is the $((k\!-\!1)N_t+\!j)^{th}$ column vector of the channel convolution matrix $\mathbf{H}$ which corresponds to $x_{k,j}$ as given by
\[
\mathbf{H} = \left[ \mathbf{h}_{1,1} \ldots  \mathbf{h}_{1,N_t}  \ldots \ldots  \mathbf{h}_{k,1} \ldots \mathbf{h}_{k,j} \ldots\mathbf{h}_{k,N_t}  \ldots  \mathbf{h}_{N,N_t}  \right]. \]
According to the WP approach, Gaussian approximation is held after the LMMSE equalization process~\cite{WangPoor1999,Tuchler2002}. In other words, the residual interference plus noise term at the output of the LMMSE equalizer can be well approximated by Gaussian distribution~\cite{WangPoor1999,Tuchler2002}. Hence, the filtered observation at time $k$ for the $j^{th}$ transmit antenna $(\hat{x}_{k,j})$ given an input symbol is assumed to have Gaussian distribution, i.e., the probability density function (pdf) of $p(\hat{x}_{k,j} \vert x_{k,j} = s) \sim CN(\mu_{k,j} s, \sigma_{k,j}^2) \text{ with } s \in S$ where $S$ is the modulation alphabet~\cite{Tuchler2002}. An equivalent model for this approximation can be written similarly to~\cite{WangPoor1999} as
\begin{align}
\label{eqn:WP_x_hat}
\hat{x}_{k,j} = \mu_{k,j} \: x_{k,j} + \eta_{k,j}, \quad k=1,\ldots,N \:\: j=1,\ldots,N_t
\end{align}
where $\eta_{k,j} \sim CN(0, \sigma_{k,j}^2)$. To reach the extrinsic information similar to~\cite{Tuchler2002}, we rearrange the expression of the filtered observation at time $k$ for the $j^{th}$ transmit antenna by setting $m_{x_{k,j}}^{prio}=0$ and $v_{x_{k,j}}^{prio}=1$ so that it does not depend on the current \textit{a priori} information $(m_{x_{k,j}}^{prio},v_{x_{k,j}}^{prio})$, which gives
\begin{align}
\label{eqn:x_hat_long}
\hat{x}_{k,j} = \mathbf{w}_{k,j}^H (\mathbf{y} - \mathbf{H} \mathbf{m}_{\mathbf{x}}^{prio} + m_{x_{k,j}}^{prio} \,\mathbf{h}_{k,j})
\end{align}
where $\mathbf{w}_{k,j}$ is the LMMSE filter coefficient vector with length $N_r(N \! + \! J)$ for the $k^{th}$ transmitted input symbol from the $j^{th}$ antenna as expressed by
\begin{small}
\begin{align}
\label{eqn:WP_w_k}
\mathbf{w}_{k,j} = \left(\!\!N_0 \mathbf{I}_{N_r(N\!+\!J)} + \!\!\!\!\!\! \sum_{\substack{i=1, i \neq k}}^N \!\!\! v_{x_{i,j}}^{prio} \mathbf{h}_{i,j} \mathbf{h}_{i,j}^H + \mathbf{h}_{k,j} \mathbf{h}_{k,j}^H \!\!\right)^{-1} \!\!\!\!\! \mathbf{h}_{k,j}
\end{align}
\end{small}
and, $\mu_{k,j}$ and $\sigma_{k,j}$ are obtained in~\cite{Tuchler2002} as
\begin{align}
\label{eqn:mu_k}
\mu_{k,j} = \mathbf{w}_{k,j}^H \: \mathbf{h}_{k,j}, \; \sigma^2_{k,j} = \mu_{k,j} (1-\mu_{k,j}^H).
\end{align}
If the $k^{th}$ transmitted symbol from the $j^{th}$ antenna is represented by $b$ bits of $[c_{k,j}^1\; c_{k,j}^2 \: \ldots \: c_{k,j}^b]$, then the extrinsic LLR value of the $q^{th}$ bit of the $k^{th}$ symbol from the $j^{th}$ antenna is expressed by considering the Gaussian assumption in (\ref{eqn:WP_x_hat}) as
\begin{footnotesize}
\begin{align*}
L_E(c_{k,j}^q) = \ln & \left(\frac{\sum\limits_{s \in S_{q,0}} p(x_{k,j} = s \vert \hat{x}_{k,j})}{\sum\limits_{s\in S_{q,1}} p(x_{k,j} = s \vert \hat{x}_{k,j})}\right) - \ln\left(\frac{\sum\limits_{s\in S_{q,0}} p(x_{k,j} = s)}{\sum\limits_{s\in S_{q,1}} p(x_{k,j} = s)}\right) 
\end{align*}
\end{footnotesize}for $q=1,2, \ldots ,b$ where $S_{q,0} \; (S_{q,1})$ denotes the subset of the modulation alphabet $S$ with symbols whose $q^{th}$ bit is $0 \; (1)$, and $p(x_{k,j} = s)$'s are the \textit{a priori} symbol probability for the $k^{th}$ transmitted symbol from the $j^{th}$ antenna. Using Bayes Rule~\cite{Bertsekas2002}, $L_E(c_{k,j}^q)$ is rewritten by considering the Gaussian assumption in (\ref{eqn:WP_x_hat}) as
\begin{footnotesize}
\begin{align}
\label{eqn:LLR_WP} 
\nonumber
L_E(c_{k,j}^q) =& \ln\left(\frac{\sum_{s\in S_{q,0}} p(\hat{x}_{k,j} \vert x_{k,j} = s) p(x_{k,j} = s)}{\sum_{s\in S_{q,1}} p(\hat{x}_{k,j} \vert x_{k,j} = s) p(x_{k,j} = s)}\right) - \\
& \ln\left(\frac{\sum_{s\in S_{q,0}} p(x_{k,j} = s)}{\sum_{s\in S_{q,1}} p(x_{k,j} = s)}\right) \; q=1,2, \ldots ,b ;
\end{align}
\end{footnotesize}
where $p(\hat{x}_{k,j} \vert x_{k,j} = s) \propto \exp(-\vert \hat{x}_{k,j} - \mu_{k,j} s \vert ^2 / \sigma_{k,j}^2)$.

As can be seen in (\ref{eqn:x_hat_long})-(\ref{eqn:mu_k}), the complexity of finding $\hat{x}_{k,j}$, $\mu_{k,j}$ and $\sigma_{k,j}$ values is $O(N^3N_r^3)$ and mainly determined by (\ref{eqn:WP_w_k}) which involves a matrix inversion of size $N_r(N\!+\!J)$. Moreover, there is no mathematical simplification in the extrinsic bit LLR expression in (\ref{eqn:LLR_WP}) for $M$-QAM signalling due to the summation over symbols unlike the BPSK signalling case discussed in~\cite{Guo2008}. Hence, this version of WP approach is not suitable for the graph based LMMSE equalization. The expressions in Proposition 1 below provide the key connection between the graph outputs (\textit{a posteriori} mean and variance values) and the WP parameters ($\hat{x}_{k,j}$, $\mu_{k,j}$ and $\sigma_{k,j}$) with no major complexity increase.

\noindent\textbf{{Proposition 2:}} WP parameters necessary to evaluate the extrinsic LLR can be found based on the  graph outputs, namely a posteriori mean and variance values, through expressions
\begin{align}
\label{eqn:x_hat_prop}
\hat{x}_{k,j} =& \left(\frac{m_{x_{k,j}}^{post}}{v_{x_{k,j}}^{post}} - \frac{m_{x_{k,j}}^{prio}}{v_{x_{k,j}}^{prio}} \right) / \left(1 + \frac{1}{v_{x_{k,j}}^{post}} - \frac{1}{v_{x_{k,j}}^{prio}} \right), \\
\label{eqn:mu_sigma_prop}
\frac{\mu_{k,j}}{\sigma^2_{k,j}} =& \left(1 + \frac{1}{v_{x_{k,j}}^{post}} - \frac{1}{v_{x_{k,j}}^{prio}} \right).
\end{align}
The proof of Proposition 2 is given in Appendices.

With the help of (\ref{eqn:mu_k}) and (\ref{eqn:x_hat_prop}-\ref{eqn:mu_sigma_prop}), the parameters of WP method ($\hat{x}_{k,j}$, $\mu_{k,j}$, $\sigma_{k,j}$) are easily computed by applying simple operations to the graph outputs and utilized in (\ref{eqn:LLR_WP}) to reach the extrinsic bit LLRs related to each transmitted symbol. When we consider the computational complexity of the proposed extrinsic bit LLR computation algorithm, the dominant contribution is due to (\ref{eqn:LLR_WP}) which has $O(N N_tM \log_2M)$ complexity per turbo iteration. To reach the overall complexity of the presented turbo receiver structure, one may consider this part, too. However, any equalizer structure using $M$-QAM modulation requires an algorithm to obtain the bit LLR values from the symbol probabilities which results in a complexity similar to that of (\ref{eqn:LLR_WP}).

\subsection{Simulation Results for the Simplified WP Approach}
\label{sec:Simulation_LLR}
\begin{figure}[t]
   \centering
   \includegraphics[width=0.47\textwidth]{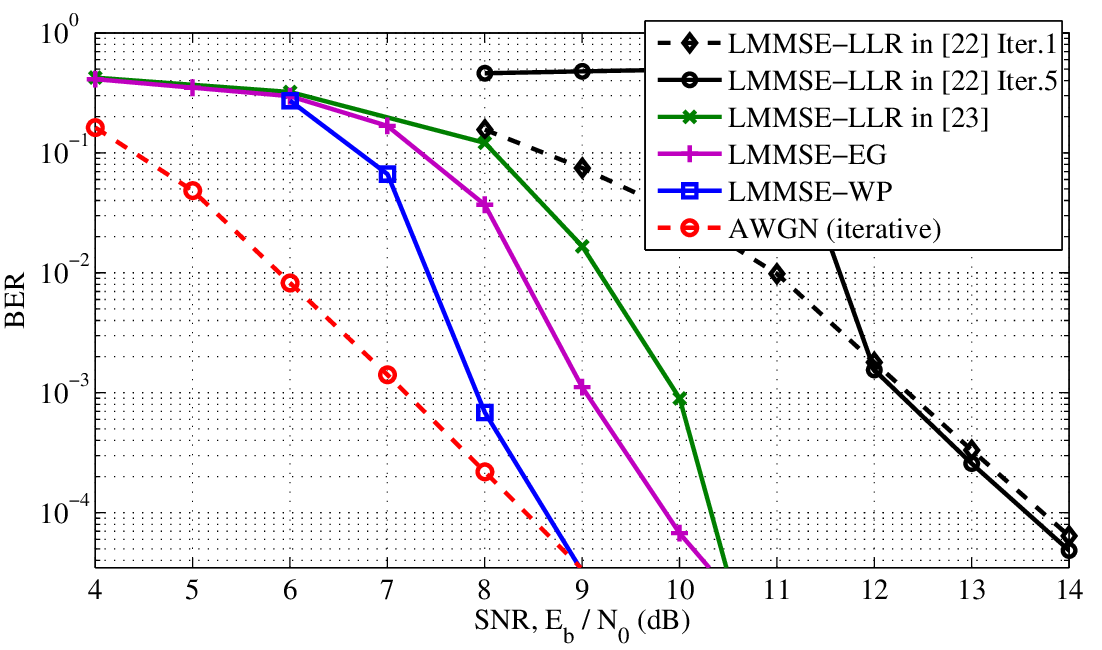}
   \caption{Performance Comparison of Extrinsic LLR Computation Algorithms for LMMSE Equalizer ($64$-QAM with ISI Channel of $\frac{1}{\sqrt{6}}\left[1 \: 2 \: 0 \: 0 \: 0 \: 1 \right]$)}
   \label{fig:llr_comp}
\end{figure}

The performance results of our proposed extrinsic bit LLR computation method, which is called LMMSE-WP, for $64$-QAM signalling as compared to the ones in~\cite{Guo2011,Pinar2014, Springer2011} are given in Fig.~\ref{fig:llr_comp}. Simulations are conducted for a SISO system under the static ISI channel whose tap amplitudes are given by $\overline{\mathbf{h}}=\frac{1}{\sqrt{6}} \left[1 \: 2 \: 0 \: 0 \: 0 \: 1 \right]$. A convolutional code with rate $1/2$ and generator polynomial $(133,171)$ is used, the data length is set to be $1800$ uncoded bits, and $5$ turbo iterations are conducted.

To serve as a benchmark for the performances in ISI channel, we also simulate the LMMSE equalizer under $5$ turbo iterations for AWGN (single-tap) channel shown by red dashed line called AWGN (iterative) to let the AWGN performance to be improved by turbo iterations under the bit interleaved coded modulation with large signaling space ($64$-QAM)~\cite{BICM1,BICM2}. 

Among the LMMSE equalizer performances, it is seen that at $10^{-4}$ BER level, there is more than $4$ dB and nearly $2$ dB gain of the proposed method with respect to the LLR exchange schemes in~\cite{Guo2011} and~\cite{Pinar2014} respectively. Another important point to mention is that the method in~\cite{Guo2011} given by equation $(8)$ leads to no improvement in performance as the number of turbo iterations increases. Also, our previous heuristic method described in~\cite{Pinar2014} needs a scaling operation which multiplies the bit LLR values at the output of the LMMSE equalizer to reach the presented performance in Fig.~\ref{fig:llr_comp}. Since finding the optimal scalar value requires exhaustive search for each different configuration, the method in~\cite{Pinar2014} is not a practical solution either. Another method is to compute the extrinsic information (in terms of mean and variance) in Gaussian domain and obtain the extrinsic LLRs using this information~\cite{Springer2011}. We call this as LMMSE-EG in Fig.~\ref{fig:llr_comp}, which causes a performance loss of $1$ dB as compared to our method. Since LMMSE-EG has a closer performance to our LMMSE-WP than the others, we continue to observe its behavior in other simulations as well. Overall, for this scenario, the simplified version of the WP approach for factor graphs is the best choice for $M$-QAM signaling among the other proposed solutions. Hence, we use this method for the LLR computation in the rest of our study. 

\begin{figure}[t]
   \centering
   \includegraphics[width=0.475\textwidth]{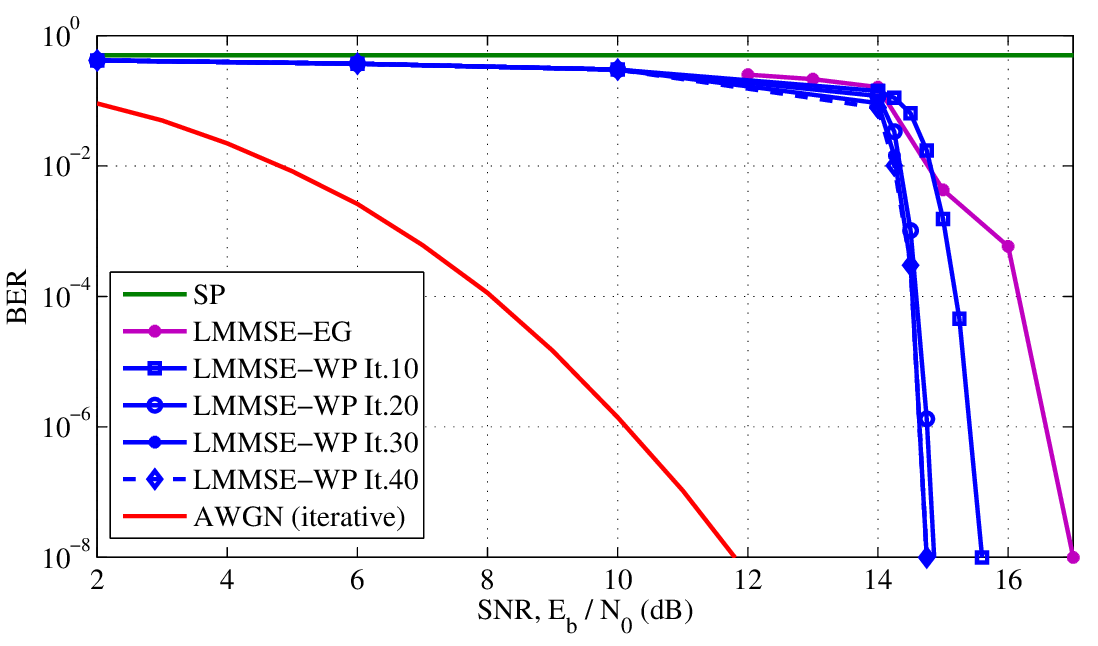}
   \caption{Performance of Various Algorithms under Severe ISI Channel of $[0.227, 0.460, 0.688,0.460, 0.227]$ ($16$-QAM)}
   \label{fig:severe_isi}
\end{figure}

We next consider a severely distorted $5$-tap ISI channel taken from~\cite{Guo2008} with coefficients $[0.227, 0.460, 0.688,0.460, 0.227]$ under $16$-QAM modulation. A convolutional code with rate $1/2$ and generator polynomial $(5,7)$ is used and the data length is set to be $40000$ uncoded bits. The performance of LMMSE-WP for different number of turbo iterations is shown in Fig.~\ref{fig:severe_isi} as compared to the benchmark AWGN performance. We also simulate LMMSE-EG, but its BER does not monotonically decreases with increasing number of turbo iterations. Thus, its best performance for each SNR value is plotted in Fig.~\ref{fig:severe_isi}. Since the other methods for LMMSE equalization mentioned above are much worse than these two, they are not included. It is seen that although suboptimality of LMMSE results in a performance gap to the benchmark, there is a sharp improvement in the performance of LMMSE-WP around $14$ dB which has a $3$ dB gain as compared to the LMMSE-EG. Moreover, sum product algorithm~\cite{Colavolpe2005} does not converge in this case. We should note that the late but sharp improvement of LMMSE-WP is not surprising because this severe ISI channel is simulated for BPSK in~\cite{Guo2008} where BCJR and LMMSE equalizers converges to the benchmark at BER values of $10^{-4}$ and $10^{-6}$ respectively. Here, the performance gap is increased due to larger modulation size.  

\begin{figure}[t]
   \centering
   \includegraphics[width=0.475\textwidth]{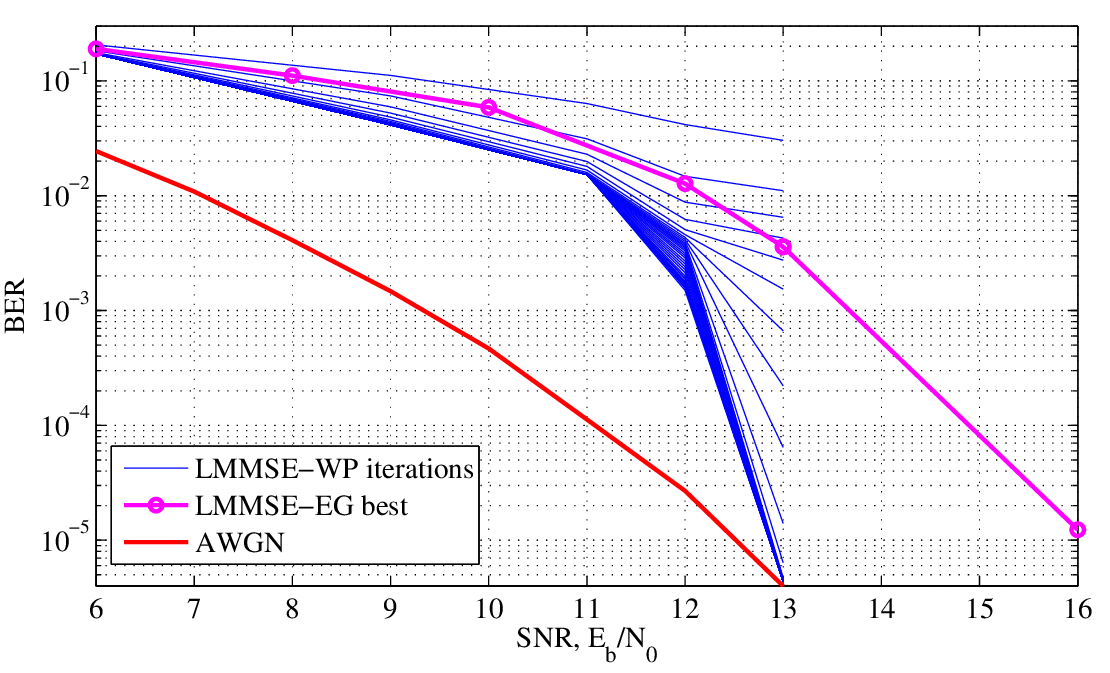}
   \caption{Performance of LMMSE-WP for $10$ Randomly Generated Channels ($64$-QAM)}
   \label{fig:10_channel}
\end{figure}

To further observe the convergence behavior of LMMSE-WP, we provide its performance results for turbo iterations $1$ to $20$ with $64$-QAM modulation over $10$ randomly generated ISI channels with $5$ taps similar to~\cite{Guo2008} in Fig.~\ref{fig:10_channel}. Energy of each ISI channel is normalized to $1$ and coefficients of each channel is randomly and independently chosen from Rayleigh distribution. A convolutional code with rate $1/2$ and generator polynomial $(5,7)$ is used and the data length is set to $36000$ bits. To emphasize the difference between LLR computation methods, the best performance of LMMSE-EG is included in Fig.~\ref{fig:10_channel}. We can see from the results that LMMSE-WP converges around $10$ turbo iterations, whereas LMMSE-EG has an error floor and results in a loss of $3$ dB at a BER of $10^{-5}$. Although we are still in search of a better way to analyze this convergence behavior over all channel conditions, Fig.~\ref{fig:10_channel} shows that LMMSE equalizer with the proposed LLR computation method is the best among all the others in the literature for various type of channels and modulation orders. 

\subsection{Convergence Properties of the Proposed Receiver}
\begin{figure}[t]
   \centering
   \includegraphics[width=0.475\textwidth]{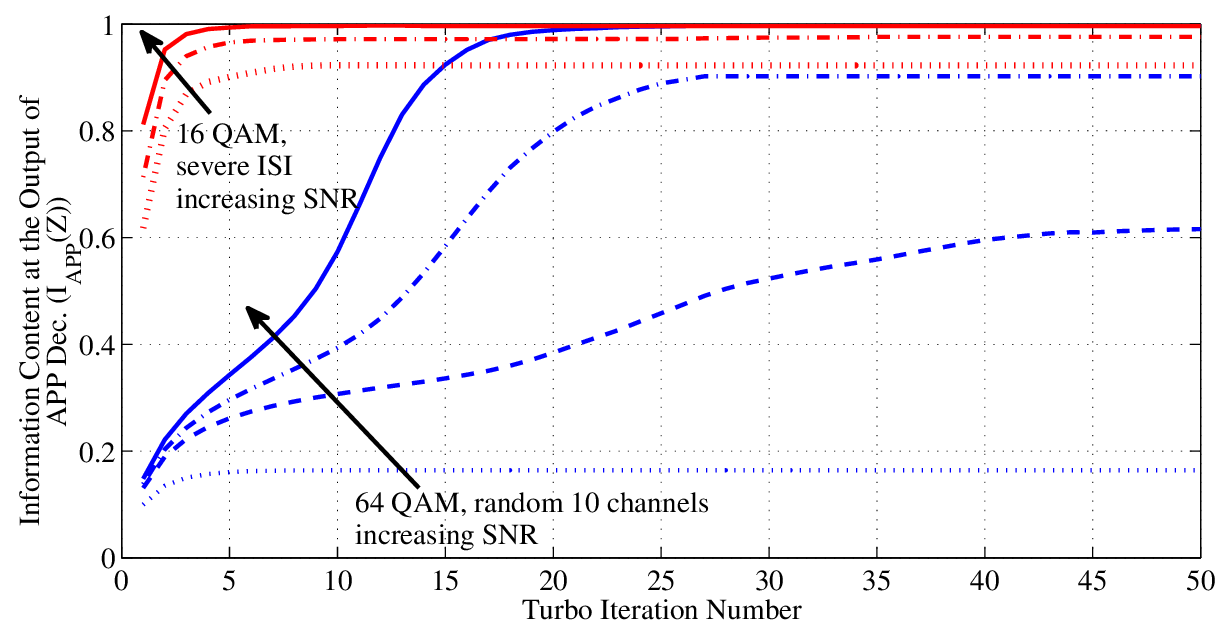}
   \caption{Convergence Characteristics of the Proposed LLR Exchange Algorithm ($16$-QAM with Severe ISI Channel of $[0.227, 0.460, 0.688,0.460, 0.227]$ and $64$-QAM with $10$ Randomly Generated ISI Channels)}
   \label{fig:llr_severe_10}
\end{figure}

In this section, the convergence properties of the proposed LLR exchange algorithm for the LMMSE equalizer are investigated similar to~\cite{TuchlerLLR,Divsalar2001,Brink2001}. Let \textit{the information content function} for the MIMO system in (\ref{eqn:observation_for_LMMSE}) is written as given in~\cite{TuchlerLLR,Divsalar2001}
\begin{small}
\begin{align}
\label{eqn:info_content}
I(Z) = \frac{1}{N N_t b} \sum_{k=1}^N \sum_{j=1}^{N_t} \sum_{q=1}^b \left[ 1-\text{log}_2(1+\exp(-Z_{k,j}^q)) \right],
\end{align}
\end{small}where $Z=\{Z_{k,j}^q\}$ is the extrinsic information sequence in log domain  for all the bits sent over the transmit antennas and its elements are expressed as
\begin{align}
\label{eqn:inform_seq_Z}
Z_{k,j}^q = (-1)^{c_{k,j}^q} L(c_{k,j}^q).
\end{align}
$L(c_{k,j}^q)$ in (\ref{eqn:inform_seq_Z}) denotes the extrinsic bit LLR value related to the $q^{th}$ bit of the $k^{th}$ transmitted symbol from the $j^{th}$ antenna. Since we are interested in the reliability at the output of the APP decoder after each turbo iteration, we compute (\ref{eqn:info_content}) by taking $L(c_{k,j}^q) = L_{APP}(c_{k,j}^q)$ in (\ref{eqn:inform_seq_Z}) where $L_{APP}(c_{k,j}^q)$ denotes the extrinsic bit LLR values at the output of the APP decoder. Using Monte Carlo simulations for the scenarios in Section~\ref{sec:Simulation_LLR}, we obtain the average information content at the output of the APP decoder with respect to the number of turbo iterations for different SNR values. As more interesting, convergence characteristics of the proposed LLR exchange method under the severe ISI scenarios is given in Fig.~\ref{fig:llr_severe_10}.

It can be seen from the results that the reliability of the extrinsic information converges with the increasing number of turbo iterations. Moreover, for larger SNR values, the information content converges to a larger value (meaning more reliable estimation) much faster. We also observe that the convergence speed is dependent on the channel condition: more severe ISI channel needs more turbo iterations. However, LMMSE-WP converges in all cases, even in severe ISI channel. 

Overall, LMMSE equalization, which is known to be advantageous in terms of its reduced complexity, superior performance results and satisfactory convergence properties for BPSK signalling~\cite{Guo2008}, becomes a good solution for $M$-QAM modulation with the proposed extrinsic LLR exchange method.

\section{COMPLEXITY ANALYSIS}
\label{sec:complexity}
The major contribution to the complexity of the proposed graph structure is caused by the matrix inversions in (\ref{eqn:overall_v}-\ref{eqn:overall_m}), \eqref{composite_defnB} and \eqref{composite_defnC}. In each building block, \eqref{composite_defnB} and \eqref{composite_defnC} need to be calculated with a complexity of $O(N_r^3)$ since they involve matrix inversions of size $N_r$ thanks to the applied matrix inversion lemma. On the other hand, (\ref{eqn:overall_v}) and (\ref{eqn:overall_m}) are applied only once for every $L$ building blocks with a complexity of $O(N_t^3 L^3)$ owing to the shifting property of the state vectors as observed in (\ref{eqn:mv_shift}). Hence, it corresponds to $O(N_t^3 L^2)$ for each building block, i.e., each time instant, where there are $N$ building blocks in our system. Therefore, the overall complexity is $O(N \cdot max\{N_r^3,N_t^3 L^2\})$ which is equal to $O(N N_t^3 L^2)$ in most of the cases. As a result of this discussion, the overall complexity per symbol per transmit antenna is $O(N_t^2 L^2)$. To reach the bit level complexity, we need to add $O(N N_tM \log_2M)$ complexity of bit LLR computation method described in Section~\ref{sec:llr_exchange}. However, to make a fair comparison to the previous studies, we continue with the complexity for symbol level, i.e., $O(N N_t^3 L^2)$ since they gave their complexity analysis in this form. 

When we consider other methods in the literature,~\cite{Duman2007} proposed the belief propagation over factor graphs for frequency selective MIMO systems with a complexity of $O(NM^{N_t \tilde{L}})$ where $\tilde{L}$ is the number of non-zero channel taps. When a high order modulation alphabet is used in a dense channel, $O(NM^{N_t \tilde{L}})$ is much greater than $O(N N_t^3 L^2)$ complexity of our method. Another study in~\cite{Som2011} discussed a Markov random field based graphical model resulting in a complexity of $O(N^2 n_t^2)$ per symbol which increases proportional to the square of block length. In addition, the proposed Kalman filtering solution in~\cite{Duman20072}, which is deprived from the improvement of backward recursion (Kalman smoothing), has a complexity of $O(N N_t^3 L^3)$. Also, the complexity of the lately studied LMMSE equalizer in~\cite{Springer2011} which was proposed to implement using a different factor graph structure from ours is $O(N N_t^3 L^3)$ which is still greater than the complexity of the structure in this study. Moreover, although the result of the LMMSE estimation in~\cite{Springer2011} is the same as our graph output on Gaussian domain, it results in an error floor for large SNR values due to their LLR exchange algorithm between Gaussian and binary domains. Overall, our proposed LMMSE solution is a practical receiver for high data rate applications with its lower complexity than those presented in the literature and its close performance to matched filter bound to be presented in the subsequent section.

\section{SIMULATION RESULTS}
\label{sec:Simulation}
We conduct our simulations under quasi-static Rayleigh fading channels with independent ISI taps, i.e., each tap is constant over one block and change independently from block to block. The ISI channel between each transmit-receive antenna pair has identical, equal power delay profile similar to the studies in~\cite{Springer2011,Springer2012,Yuan2008}, i.e., all $L$ taps have equal power which is normalized so that the total power of channel response is unity $ \sum_{k=0}^{L-1} E\{ {\vert h_{ij}(k)\vert}^2 \} = 1$, where $h_{ij}(k)$ is the $k^{th}$ channel tap between the $j^{th}$ transmit antenna and $i^{th}$ receive antenna. The simulations are based on the system model in Fig.~\ref{fig:system_model_MIMO} with a random interleaver and a rate $1/2$ convolution code whose generator matrix is $(7,5)_8$ under different modulation order. In all simulations, data bits are coded, interleaved and then modulated. The modulated symbols are distributed to the transmit antennas by a spatial multiplexing operation as given in Fig.~\ref{fig:system_model_MIMO}. 
 
For the LLR exchange process between the LMMSE equalizer and the APP decoder, we use the WP approach explained in Section~\ref{sec:llr_exchange}. With our proposed bit LLR exchange algorithm, there is no need to apply scaling operations to the extrinsic LLR values at the output of the LMMSE equalizer and the APP decoder to improve the performance contrary to the turbo decoding algorithms in the literature~\cite{Colavolpe2011,Springer2012},~\cite{Anderson2013,Colavolpe2007}. 

For all the configurations below, we also provide the matched filter bound (MFB) performances as a benchmark to make a comparison. The MFB performances are obtained under the assumption that the symbols which cause interference to the interested symbol due to multi-path and multi-antenna effects are perfectly known by the receiver for each interested symbol~\cite{Barry2004}. Hence, it is practically impossible to reach MFB performance for any receiver structure. We take MFB performance as a genie-aided lower bound for the proposed scheme.

\begin{figure}[t]
   \centering
   \includegraphics[width=0.475\textwidth]{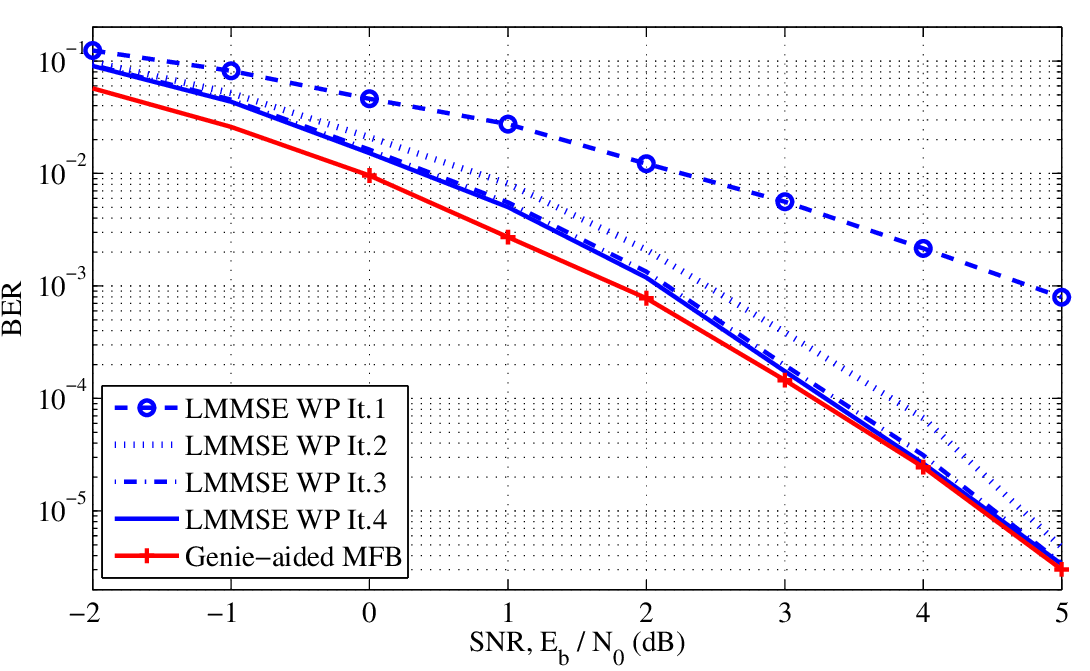}
   \caption{BER Performance of the Factor Graph Based LMMSE Equalizer ($2\times 2$ MIMO with $L=5$ for BPSK Signaling)}
   \label{fig:LMMSE_FG_MIMO_BPSK}
\end{figure}

BER performance of the proposed factor graph based LMMSE equalizer is given in Fig.~\ref{fig:LMMSE_FG_MIMO_BPSK} for BPSK signalling with $N_t=N_r=2$ under a $5$-tap channel. The data length is set to $4096$ bits. This is the same configuration as the one in~\cite{Springer2011} except the interleaver type which is S-random in~\cite{Springer2011}. For BPSK signaling, all of the llr exchange algorithms mentioned in Section~\ref{sec:Simulation_LLR} (with a clear modification to real transmission) are reduced to the same simple expression given in Proposition $1$ in~\cite{Guo2008}. Thus, the proposed method must have identical performance with the one in~\cite{Springer2011} when using the same interleavers since both algorithms implement time domain LMMSE filtering operation. However, there is an error floor observed in high SNR regions in~\cite{Springer2011}, which is caused by the llr exchange algorithm that is not modified according to real transmission. On the other hand, the performance of the proposed method is very close to the MFB below the BER value of $10^{-3}$ without any diversity loss or error floor, which can be obtained by using the other llr computation methods modified according to real transmission. From Fig.~\ref{fig:LMMSE_FG_MIMO_BPSK}, it is seen that only $3$ iterations are sufficient for LMMSE equalizer under this configuration. Thus, similar to the SISO case, LMMSE is a good solution to MIMO ISI equalization for BPSK signaling.

\begin{figure}[t]
   \centering
   \includegraphics[width=0.475\textwidth]{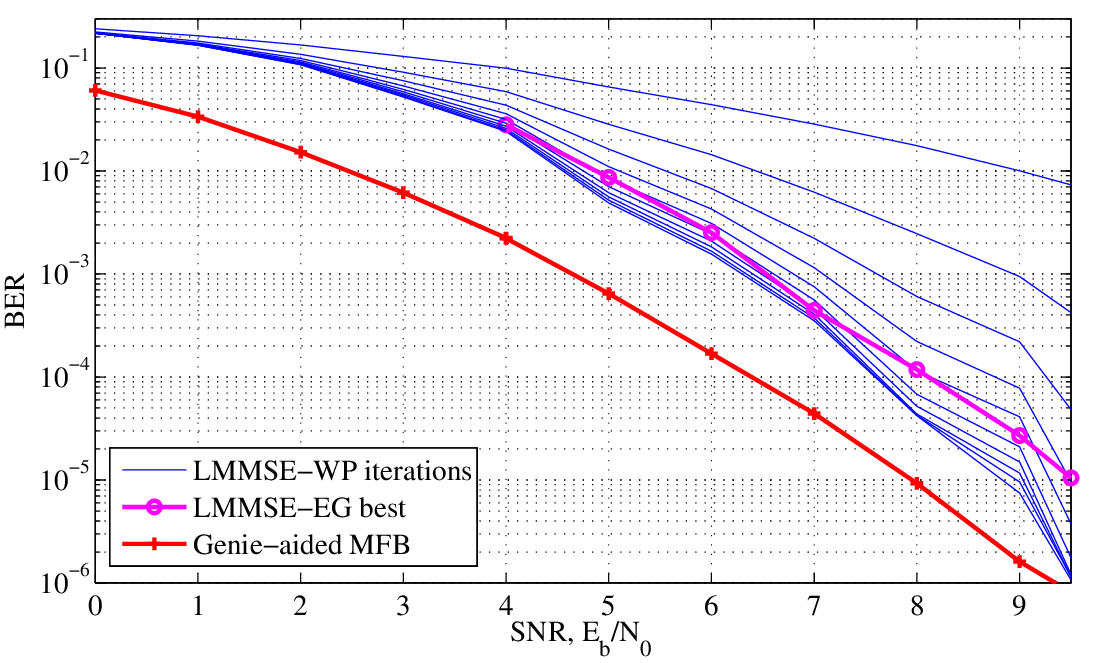}
   \caption{BER Performance of the Factor Graph Based LMMSE Equalizer ($2\times 2$ MIMO with $L=4$ for 16-QAM Signaling)}
   \label{fig:LMMSE_FG_MIMO_16QAM}
\end{figure} 
\begin{figure}[t]
   \centering
   \includegraphics[width=0.475\textwidth]{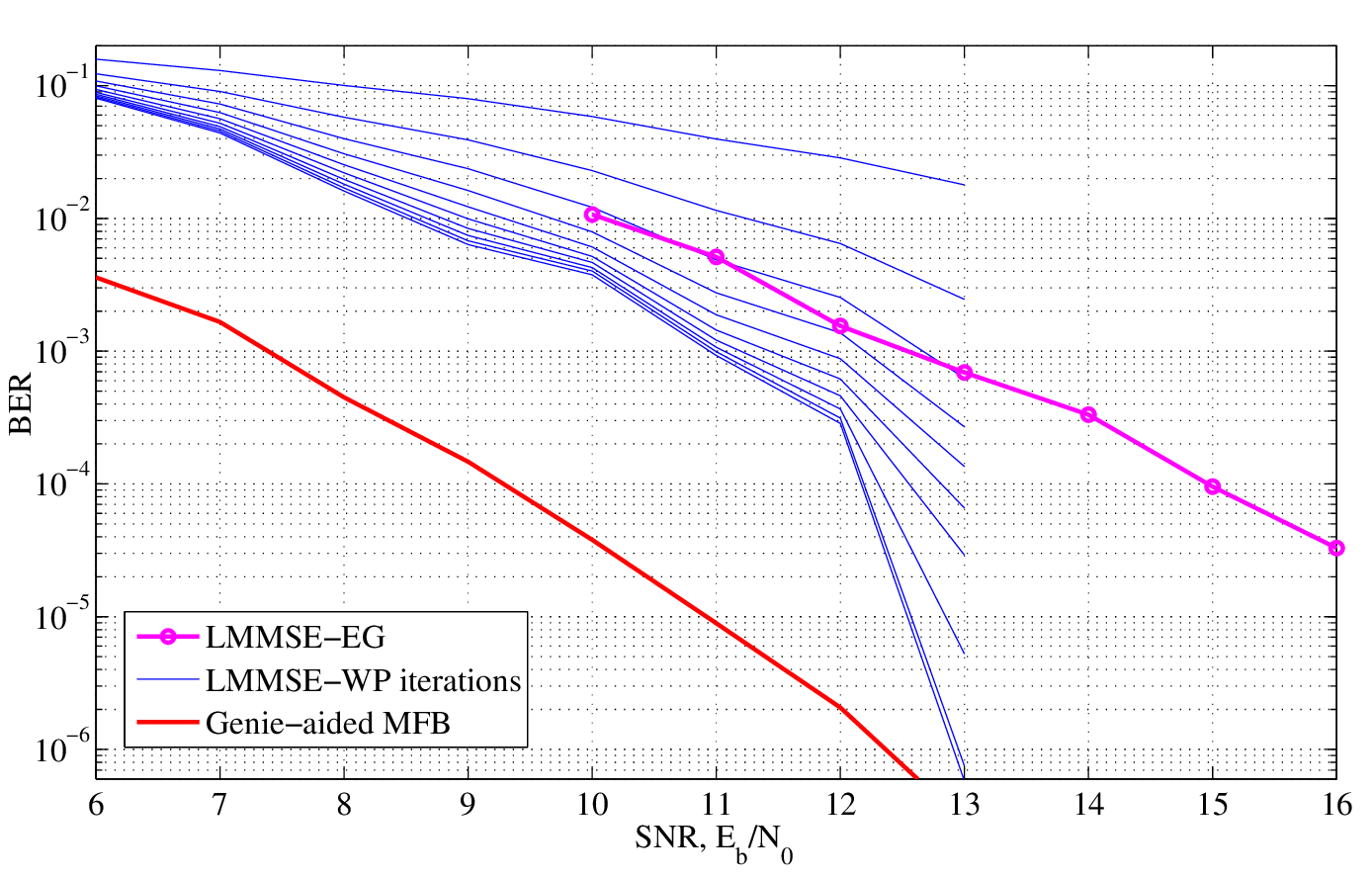}
   \caption{BER Performance of the Factor Graph Based LMMSE Equalizer ($2\times 2$ MIMO with $L=4$ for 64-QAM Signaling)}
   \label{fig:LMMSE_FG_MIMO_64QAM}
\end{figure}

The important performance difference between our proposed llr computation method and the others is observed for larger constellations, which is also presented in Section~\ref{sec:Simulation_LLR}. Hence, we would like to present the performance of such a challenging scenario with higher order constellations for MIMO ISI transmission this time. Fig.~\ref{fig:LMMSE_FG_MIMO_16QAM} depicts simulation results for $16$-QAM signaling under a $4$-tap ISI channel with $N_t=N_r=2$. The data length is set to $4096$ bits. It can be seen from Fig.~\ref{fig:LMMSE_FG_MIMO_16QAM} that the proposed method has a performance which is less than $1$ dB away from the MFB performance below the BER value of $10^{-4}$ for $7$ turbo iterations. The increased constellation size leads to a higher number of turbo iterations for good performance, but turbo iteration number is not a direct multiplier of computational complexity since all packets do not require $7$ iterations. Moreover, the constellation size $M$ is included only in the complexity term related to the bit LLR computation method in a linearly increasing fashion. Hence, our method is a practical choice as a receiver structure with its solid performance while achieving higher data rates. However, LMMSE-EG, the performance of the method in~\cite{Springer2011}, suffers from an error floor in high SNR region; although, it is represented by its best performance among $20$ turbo iterations. This difference is observed more dramatically in the next result where $64$-QAM is used.

Fig.~\ref{fig:LMMSE_FG_MIMO_64QAM} presents simulation result for $64$-QAM signaling under a $4$-tap ISI channel with $N_t=N_r=2$. The data length is set to $12000$ bits. Although there is a performance gap of LMMSE-WP to the MFB, it converges sharply to this benchmark after BER value of $10^{-4}$. Moreover, $10$ turbo iterations are sufficient for this performance. However, LMMSE-EG, which is represented by its best performance among $20$ turbo iterations, has an error floor resulting in more than $3$ dB loss for high SNR region. Thus, LMMSE-WP is superior than LMMSE-EG in terms of computational complexity and performance for MIMO transmission as well. 

Asymmetric MIMO case such as $4 \times 6$ with $4$ tap ISI channel with $64$-QAM modulation is also simulated where LMMSE equalizer converges much faster than the case in Fig.~\ref{fig:LMMSE_FG_MIMO_64QAM}. Moreover, other scenarios including larger rate convolutional codes and/or S-random interleavers designed similar to~\cite{heegard_turbo, Tugcan_tez} are simulated. It is observed that using S-random interleaver for Rayleigh block fading MIMO ISI channels with $M$-QAM modulations does not provide any significant improvement. Moreover, although increasing rate of convolutional code results in later convergence to the genie-aided MFB, performance of LMMSE-WP eventually gets very close or almost identical to that lower bound for large SNR values. 

Consequently, all these comparisons show that LMMSE-WP is a good solution for also MIMO ISI channel with the proposed llr exchange method and reduced complexity.

\section{CONCLUSION}
\label{sec:conclsn}

In this study, we developed a factor graph structure for the LMMSE equalization of frequency selective MIMO channels. Our proposed graph has the advantage of low complexity as compared to the conventional block LMMSE filtering operation and the other graph based LMMSE filtering approaches in the literature. In addition, we provided an efficient way of computing extrinsic LLR values of LMMSE equalization for $M$-QAM constellations based on the well-known Wang-Poor (WP) approach with no major complexity increase. In other words, we have shown the mathematical relation between the output of the LMMSE equalizer and the WP parameters in a suitable fashion for factor graph. To sum up, we proposed a low complexity, practical LMMSE equalizer for turbo decoding of MIMO ISI channels with a good performance as confirmed by our simulation results. Our method comes forefront particularly for higher constellation sizes with its low computational complexity owing to the Gaussian assumption used in the factor graph and the proposed bit LLR exchange algorithm.

\section*{APPENDICES}
\noindent\textbf{Proof of Proposition 1:} In this section, the proof of the equivalency of $\left(\mathbf{V}_{\overline{\mathbf{x}}_k}^{post}, \mathbf{m}_{\overline{\mathbf{x}}_k}^{post}\right)$ in this paper and $\left(\mathbf{\Sigma}_{11}^{(k)^{-1}}, \mathbf{\mu}_{11}^{(k)}\right)$ in~\cite{Springer2011} is given for the steady state. One can easily show the equivalency for the transient states following the same steps. In the following steps, $(\mathbf{\Sigma}, \mathbf{\mu})$ denotes the messages in~\cite{Springer2011} with the corresponding indices.

\noindent\textbf{{Claim 1:}} $\overrightarrow{\mathbf{V}}_{\overline{\mathbf{x}}_k^{''}} = \mathbf{\Sigma}_{6}^{(k)}$ and $\overleftarrow{\mathbf{W}}_{\overline{\mathbf{x}}_k^{''}} = \mathbf{\Sigma}_{5}^{(k)^{-1}}$.

It follows from Claim 1 that $\mathbf{V}_{\overline{\mathbf{x}}_k}^{post} = \mathbf{\Sigma}_{11}^{(k)^{-1}}$.

\noindent\textbf{{Proof of Claim 1:}} Using proof by induction method, we can obtain $\overrightarrow{\mathbf{V}}_{\overline{\mathbf{x}}_k^{''}}^{-1}$ along the way of forward recursion from $\overrightarrow{\mathbf{V}}_{\overline{\mathbf{x}}_{k-1}^{''}}^{-1}$ by the message passing rules in Table~\ref{tab:table1} as follows
\begin{align*}
\overrightarrow{\mathbf{V}}_{\overline{\mathbf{x}}_k^{''}}^{-1} = \overline{\mathbf{H}}^H \mathbf{V}_{\mathbf{n}}^{-1} \overline{\mathbf{H}} +  \left[ \begin{array}{cc} 
[\mathbf{\Sigma}_{6}^{(k-1)}]_{N_t+1:L N_t}^{-1} & \mathbf{0} \\
\mathbf{0} & \mathbf{V}_{\mathbf{x}_{k}}^{\downarrow}
\end{array} \right] = \mathbf{\Sigma}_{6}^{(k)^{-1}},
\end{align*}
where $\mathbf{V}_{\mathbf{n}}$ is the covariance matrix of the noise vector $\mathbf{n}_k$, $[A]_{a:b}$ denotes the sub-matrix of $A$ composed of the elements located between rows $a:b$ and columns $a:b$, and $[A]_{a:b}^{-1}$ denotes the inverse of $[A]_{a:b}$.

To prove $\overleftarrow{\mathbf{W}}_{\overline{\mathbf{x}}_k^{''}} = \mathbf{\Sigma}_{5}^{(k)^{-1}}$, it suffices to show that \[[\overleftarrow{\mathbf{W}}_{\overline{\mathbf{z}}_{k+1}}]_{1:J N_t} = [\mathbf{\Sigma}_{7}^{(k+1)}]_{1:J N_t}^{-1}.\] Along the way of backward recursion from $\overleftarrow{\mathbf{W}}_{\overline{\mathbf{z}}_{k+2}}$ using Table~\ref{tab:table1}, we have
\begin{align*}
\overleftarrow{\mathbf{W}}_{\overline{\mathbf{x}}_{k+1}} = &  \overline{\mathbf{H}}^H \mathbf{V}_{\mathbf{n}}^{-1} \overline{\mathbf{H}} +  \left[ \begin{array}{cc} 
\mathbf{0} & \mathbf{0} \\
\mathbf{0} & [\overleftarrow{\mathbf{W}}_{\overline{\mathbf{z}}_{k+2}}]_{1:J N_t}
\end{array} \right] \\
\triangleq & K \triangleq \left[ \begin{array}{cc} 
K_1 & K_2  \\
K_3 & (K_4)_{N_t \times N_t}
\end{array} \right], \quad \text{and}
\end{align*}
\begin{align*}
\overleftarrow{\mathbf{W}}_{\overline{\mathbf{z}}_{k+1}} = & \left( \overleftarrow{\mathbf{V}}_{\overline{\mathbf{x}}_{k+1}} + \left[ \begin{array}{cc} 
\mathbf{0} & \mathbf{0} \\
\mathbf{0} & \mathbf{V}_{\mathbf{x}_{k+1}}^{\downarrow} 
\end{array} \right] \right)^{-1} \\
= & \left[ \begin{array}{cc} 
\mathbf{I} & K_2 \mathbf{V}_{\mathbf{x}_{k+1}}^{\downarrow}   \\
\mathbf{0} & \mathbf{I} + K_4  \mathbf{V}_{\mathbf{x}_{k+1}}^{\downarrow} 
\end{array} \right]^{-1} \cdot K,
\end{align*}
where inverse of a block diagonal matrix in~\cite{Petersen2012} gives
\begin{align*}
[\overleftarrow{\mathbf{W}}_{\overline{\mathbf{z}}_{k+1}}]_{1:J N_t} = K_1 - K_2  \left(\mathbf{V}_{\mathbf{x}_{k+1}}^{\downarrow -1} + K_4 \right)^{-1} K_3.
\end{align*}
Also, by induction method, it can be shown that
\begin{align}
\label{eqn:sigma7}
\mathbf{\Sigma}_{7}^{(k+1)} = & \left(K + \left[ \begin{array}{cc} 
\mathbf{0} & \mathbf{0} \\
\mathbf{0} & \mathbf{V}_{\mathbf{x}_{k+1}}^{\downarrow -1} 
\end{array} \right] \right)^{-1}.
\end{align}
And, the result follows from the inversion rule of block diagonal matrix~\cite{Petersen2012} as given by
\begin{align*}
[\mathbf{\Sigma}_{7}^{(k+1)}]_{1:J N_t} = \left( K_1 - K_2  \left(\mathbf{V}_{\mathbf{x}_{k+1}}^{\downarrow -1} + K_4 \right)^{-1} K_3 \right)^{-1}.
\end{align*}
It should be noted that $\overleftarrow{\mathbf{W}}_{\overline{\mathbf{z}}_{k+1}} \neq \mathbf{\Sigma}_{7}^{(k+1)^{-1}}$. On the other hand, what we have proved is that $ [\overleftarrow{\mathbf{W}}_{\overline{\mathbf{z}}_{k+1}}]_{1:J N_t} = [\mathbf{\Sigma}_{7}^{(k+1)}]_{1:J N_t}^{-1}$.

\noindent\textbf{{Claim 2:}} $\overrightarrow{\mathbf{W}}_{\overline{\mathbf{x}}_k^{''}} \overrightarrow{\mathbf{m}}_{\overline{\mathbf{x}}_k^{''}} \!\! =  \mathbf{\Sigma}_{6}^{(k)^{-1}} \!\! \mathbf{\mu}_{6}^{(k)},\overleftarrow{\mathbf{W}}_{\overline{\mathbf{x}}_k^{''}}  \overleftarrow{\mathbf{m}}_{\overline{\mathbf{x}}_k^{''}} = \mathbf{\Sigma}_{5}^{(k)^{-1}} \!\! \mathbf{\mu}_{5}^{(k)^{-1}}$.

It follows from Claim 1 and 2 that $\mathbf{m}_{\overline{\mathbf{x}}_k}^{post} = \mathbf{\mu}_{11}^{(k)^{-1}}$.

\noindent\textbf{{Proof of Claim 2:}} Similarly, by the induction method and the message passing rules in Table~\ref{tab:table1} together with the results of Claim 1, one can write
\begin{footnotesize}
\begin{align*}
\overrightarrow{\mathbf{W}}_{\overline{\mathbf{x}}_k^{''}} \overrightarrow{\mathbf{m}}_{\overline{\mathbf{x}}_k^{''}} = &  \overline{\mathbf{H}}^H \mathbf{V}_{\mathbf{n}}^{-1} {\mathbf{y}_k} +  \left[ \!\!\! \begin{array}{cc} 
[\overrightarrow{\mathbf{V}}_{\overline{\mathbf{x}}_{k-1}^{''}}]_{N_t+1:L N_t} & \!\!\! \mathbf{0} \\
\mathbf{0} & \!\!\! \mathbf{V}_{\mathbf{x}_{k}}^{\downarrow}
\end{array} \!\!\! \right]^{-1}  \cdot \\
& \quad\quad\quad\quad\quad\quad\quad\quad \left[ \!\!\! \begin{array}{c} 
[\overrightarrow{\mathbf{m}}_{\overline{\mathbf{x}}_{k-1}^{''}}]_{N_t+1:L N_t} \\
\mathbf{m}_{\mathbf{x}_{k}}^{\downarrow}
\end{array} \!\!\! \right] \\
= \overline{\mathbf{H}}^H \mathbf{V}_{\mathbf{n}}^{-1} {\mathbf{y}_k} &+  \left[  \begin{array}{c} \!\!\!\!
[\mathbf{\Sigma}_{6}^{(k-1)}]_{N_t+1:L N_t}^{-1} [\mathbf{\mu}_{6}^{(k-1)}]_{N_t+1:L N_t} \\ \\
\mathbf{V}_{\mathbf{x}_{k}}^{\downarrow -1} \; \mathbf{m}_{\mathbf{x}_{k}}^{\downarrow} 
\end{array} \!\!\!\! \right] = \mathbf{\Sigma}_{6}^{(k)^{-1}} \mathbf{\mu}_{6}^{(k)}.
\end{align*}
\end{footnotesize}
To prove $\overleftarrow{\mathbf{W}}_{\overline{\mathbf{x}}_k^{''}} \overleftarrow{\mathbf{m}}_{\overline{\mathbf{x}}_k^{''}} = \mathbf{\Sigma}_{5}^{(k)^{-1}} \mathbf{\mu}_{5}^{(k)^{-1}}$, it suffices to show that \[[\overleftarrow{\mathbf{W}}_{\overline{\mathbf{z}}_{k+1}} \; \overleftarrow{\mathbf{m}}_{\overline{\mathbf{z}}_{k+1}}]_{1:J N_t}= [\mathbf{\Sigma}_{7}^{(k+1)}]_{1:J N_t}^{-1} \: [\mathbf{\mu}_{7}^{(k+1)}]_{1:J N_t}.\] Along the way of backward recursion from $\overleftarrow{\mathbf{W}}_{\overline{\mathbf{z}}_{k+2}}$ using Table~\ref{tab:table1}, we have
\begin{align*}
\overleftarrow{\mathbf{W}}_{\overline{\mathbf{z}}_{k+1}} \overleftarrow{\mathbf{m}}_{\overline{\mathbf{z}}_{k+1}} = & \left( \mathbf{I} - K F C F^H \right) \left( \tilde{K} - K F \mathbf{m}_{\mathbf{x}_{k}}^{\downarrow}  \right), \; \text{where} \\
K \triangleq  \overleftarrow{\mathbf{W}}_{\overline{\mathbf{x}}_{k+1}}     \triangleq & \left[ \begin{array}{cc}
{K}_1 & K_2\\
{K}_3 & K_4
\end{array} \right], \; C \triangleq  \left( \mathbf{V}_{\mathbf{x}_{k+1}}^{\downarrow -1} + F^H K F \right)^{-1}\\
\tilde{K} \triangleq & \overleftarrow{\mathbf{W}}_{\overline{\mathbf{x}}_{k+1}} \overleftarrow{\mathbf{m}}_{\overline{\mathbf{x}}_{k+1}} \triangleq  \left[ 
\tilde{K}_1^T \; \tilde{K}_2^T \right]^T.
\end{align*}
By simple matrix operations one can obtain the following 
\begin{footnotesize}
\begin{align*}
[\overleftarrow{\mathbf{W}}_{\overline{\mathbf{z}}_{k+1}} \: \overleftarrow{\mathbf{m}}_{\overline{\mathbf{z}}_{k+1}}]_{1:J N_t} & = \tilde{K}_1 - \\
& K_2 \left( \mathbf{V}_{\mathbf{x}_{k+1}}^{\downarrow -1} + K_4 \right)^{-1} (\tilde{K}_2 + \mathbf{V}_{\mathbf{x}_{k+1}}^{\downarrow -1} \mathbf{m}_{\mathbf{x}_{k+1}}^{\downarrow}).
\end{align*}
\end{footnotesize}
On the other hand, using induction, $\mathbf{\mu}_7^{(k+1)}$ can be written as
\begin{align}
\label{eqn:mu7}
\mathbf{\mu}_7^{(k+1)} = & \mathbf{\Sigma}_7^{(k+1)} \left[ \begin{array}{c}
\tilde{K}_1 \\
\tilde{K}_2 + \mathbf{V}_{\mathbf{x}_{k}}^{\downarrow -1} \mathbf{m}_{\mathbf{x}_{k}}^{\downarrow}
\end{array} \right].
\end{align}
Combining (\ref{eqn:sigma7}) and (\ref{eqn:mu7}) using elementary matrix operations proves Claim 2 as follows
\begin{footnotesize}
\begin{align*}
& [\mathbf{\mu}_7^{(k+1)}]_{1:J N_t}^{-1} \: [\mathbf{\mu}_7^{(k+1)}]_{1:J N_t}  = \tilde{K}_1 + ( K_1 - K_2  (\mathbf{V}_{\mathbf{x}_{k+1}}^{\downarrow -1} + K_4 )^{-1} K_3 ) \cdot \\ 
&( -K_1^{-1} K_2 (\mathbf{V}_{\mathbf{x}_{k+1}}^{\downarrow -1} + K_4 - K_3 K_1^{-1} K_2 )^{-1} ) (\tilde{K}_2 
\mathbf{V}_{\mathbf{x}_{k+1}}^{\downarrow -1} \mathbf{m}_{\mathbf{x}_{k+1}}^{\downarrow -1} ) \\
&  =\tilde{K}_1 - K_2 \left( \mathbf{V}_{\mathbf{x}_{k+1}}^{\downarrow -1} + K_4 \right)^{-1} \cdot  (\tilde{K}_2 + \mathbf{V}_{\mathbf{x}_{k+1}}^{\downarrow -1} \mathbf{m}_{\mathbf{x}_{k+1}}^{\downarrow}).
\end{align*}
\end{footnotesize}%

\noindent\textbf{Proof of Proposition 2}: The LMMSE filter coefficient vector for the $k^{th}$ transmitted symbol from the $j^{th}$ transmit antenna, $\mathbf{w}_{k,j}$, previously given in (\ref{eqn:WP_w_k}) can be rewritten as
\begin{align}
\label{eqn:w_k_short}
\mathbf{w}_{k,j} &= \left( \mathbf{V}_{\xi_{k,j}} + \mathbf{h}_{k,j} \mathbf{h}_{k,j}^H\right)^{-1} \mathbf{h}_{k,j}, \quad \text{where} \\
\mathbf{V}_{\xi_{k,j}} & \triangleq N_0 \mathbf{I}_{N_r(N\!+\!J)} + \sum_{\substack{i=1,  i \neq k}}^N v_{x_{i,j}}^{prio} \mathbf{h}_{i,j} \mathbf{h}_{i,j}^H.
\label{eqn:v_eps_k}
\end{align}
By matrix inversion lemma~\cite{Johnson1990},  (\ref{eqn:w_k_short}) could be simplified to
\begin{align}
\label{eqn:w_k_short_mat_lemma}
\mathbf{w}_{k,j} = \frac{\mathbf{V}_{\xi_{k,j}} ^{-1} \mathbf{h}_{k,j}}{1 + \mathbf{h}_{k,j}^H \mathbf{V}_{\xi_{k,j}} ^{-1} \mathbf{h}_{k,j}}.
\end{align}
Inserting (\ref{eqn:w_k_short_mat_lemma}) into (\ref{eqn:x_hat_long}) gives
\begin{small}
\begin{align}
\label{eqn:x_hat_w_inserted}
\hat{x}_{k,j} = \frac{\mathbf{h}_{k,j}^H \mathbf{V}_{\xi_{k,j}} ^{-1}}{1 + \mathbf{h}_{k,j}^H \mathbf{V}_{\xi_{k,j}} ^{-1} \mathbf{h}_{k,j}} \left( \mathbf{y} - \mathbf{H} \mathbf{m}_\mathbf{x}^{prio} + \mathbf{h}_{k,j} m_{x_{k,j}}^{prio} \right).
\end{align}
\end{small}
The outputs of the LMMSE equalizer, the \textit{a posteriori} mean and variance values, are defined in~\cite{Steven1993} and used in~\cite{Guo2008} as
\begin{align}
\label{eqn:v_post}
v_{x_{k,j}}^{post} = \frac{1}{1/v_{x_{k,j}}^{prio} + \mathbf{h}_{k,j}^H \mathbf{V}_{\xi_{k,j}} ^{-1} \mathbf{h}_{k,j}}
\end{align}
\begin{small}
\begin{align}
\label{eqn:m_post}
m_{x_{k,j}}^{post} = \frac{ m_{x_{k,j}}^{prio} / v_{x_{k,j}}^{prio} + \mathbf{h}_{k,j}^H \mathbf{V}_{\xi_{k,j}} ^{-1} \left( \mathbf{y}- \mathbf{H} \mathbf{m}_\mathbf{x}^{prio} + \mathbf{h}_{k,j}  m_{x_{k,j}}^{prio} \right)}{1/v_{x_{k,j}}^{prio} + \mathbf{h}_{k,j}^H \mathbf{V}_{\xi_{k,j}} ^{-1} \mathbf{h}_{k,j}}.
\end{align}
\end{small} Using (\ref{eqn:x_hat_w_inserted})-(\ref{eqn:m_post}) we obtain
\begin{align}
\nonumber
 \left(\frac{m_{x_{k,j}}^{post}}{v_{x_{k,j}}^{post}} - \frac{m_{x_{k,j}}^{prio}}{v_{x_{k,j}}^{prio}}  \right) =& \mathbf{h}_{k,j}^H \mathbf{V}_{\xi_{k,j}} ^{-1} \left( \mathbf{y}- \mathbf{H} \mathbf{m}_\mathbf{x}^{prio} + \mathbf{h}_{k,j}  m_{x_{k,j}}^{prio} \right) \\ 
 \label{eqn:for_x_hat_final}
 		=& \hat{x}_{k,j} \left( 1 + \mathbf{h}_{k,j}^H \mathbf{V}_{\xi_{k,j}} ^{-1} \mathbf{h}_{k,j} \right).
\end{align}
Combining (\ref{eqn:v_post}) and (\ref{eqn:for_x_hat_final}) gives the expression for $\hat{x}_{k,j}$ in (\ref{eqn:x_hat_prop}). Note that (\ref{eqn:mu_k}) gives ${\sigma^2_{k,j}}/{\mu_{k,j}} = 1-\mathbf{h}_{k,j}^H \mathbf{w}_{k,j}$. Then, the result in (\ref{eqn:mu_sigma_prop}) follows from (\ref{eqn:w_k_short_mat_lemma}) and (\ref{eqn:v_post}).

The derivations given above provide a mathematical transition between the LMMSE equalizer outputs and the commonly used WP approach for the extrinsic LLR calculation which is very useful particularly for the graph based LMMSE algorithms for $M$-QAM modulation.

\noindent\textbf{GMP Rules}: The GMP rules for basic and composite building blocks are given in Table~\ref{tab:table1}.
\setcounter{tblEqCounter}{\theequation}
\begin{table}[t]
\caption{GMP Rules for Basic and Composite Blocks} 
\label{tab:table1}
\centering 
\renewcommand{\arraystretch}{}
\begin{tabular}{c|p{6.5cm}} 
\hline 
Blocks & GMP Rules  \\ [-0.5ex] \hline  \\ [-2.3ex]
{\includegraphics[scale=0.5]{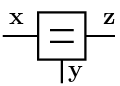}} & 
\raisebox{0.6\height}{\begin{tabular}{l c}
\footnotesize{$\overrightarrow{\mathbf{W}}_\mathbf{z} = \overrightarrow{\mathbf{W}}_\mathbf{x} + \mathbf{W}^\uparrow_\mathbf{y} \quad\quad\quad\quad\quad\quad$} & \footnotesize{\numberTblEq{equal_w}} \\ 

\footnotesize{$\overrightarrow{\mathbf{W}}_\mathbf{z} \overrightarrow{\mathbf{m}}_\mathbf{z} = \overrightarrow{\mathbf{W}}_\mathbf{x} \overrightarrow{\mathbf{m}}_\mathbf{x} + \mathbf{W}^\uparrow_\mathbf{y} \mathbf{m}^\uparrow_\mathbf{y} \;\quad\quad\quad\quad\quad\quad $} &\footnotesize{\numberTblEq{equal_m}} \end{tabular} }
\\ [-0.7ex] \hline \\ [-2.2ex]

\raisebox{-.3\height}{\includegraphics[scale=0.5]{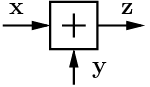}} &  
\begin{tabular}{l c}
\footnotesize{$\overrightarrow{\mathbf{V}}_\mathbf{z} = \overrightarrow{\mathbf{V}}_\mathbf{x} + \mathbf{V}^\uparrow_\mathbf{y}, \quad \overrightarrow{\mathbf{m}}_\mathbf{z} = \overrightarrow{\mathbf{m}}_\mathbf{x} + \mathbf{m}^\uparrow_\mathbf{y} \quad\quad\quad$} & \footnotesize{\numberTblEq{sum_v}} \\ 

\footnotesize{$\overleftarrow{\mathbf{V}}_\mathbf{x} = \overleftarrow{\mathbf{V}}_\mathbf{z} + \mathbf{V}^\uparrow_\mathbf{y}, \quad \overleftarrow{\mathbf{m}}_\mathbf{x} = \overleftarrow{\mathbf{m}}_\mathbf{z} - \mathbf{m}^\uparrow_\mathbf{y}\quad\quad\quad$} & \footnotesize{\numberTblEq{sum_v_back}} 

\end{tabular} 
\\ \\ [-2.5ex] \hline \\ [-2.3ex] 

\raisebox{-.3\height}{\includegraphics[scale=0.5]{figures/basic_block_matf}} & 
\begin{tabular}{l c} 
\footnotesize{$\overrightarrow{\mathbf{V}}_\mathbf{y} = \mathbf{A}\,\overrightarrow{\mathbf{V}}_\mathbf{x} \, \mathbf{A}^H, \quad \overrightarrow{\mathbf{m}}_\mathbf{y} = \mathbf{A}\,\overrightarrow{\mathbf{m}}_\mathbf{x} \;\:\quad \quad\quad\quad$} & \footnotesize{\numberTblEq{matf_v}}
\end{tabular}
\\ \\ [-2.3ex] \hline \\ [-2.3ex] 

\raisebox{-.2\height}{\includegraphics[scale=0.5]{figures/basic_block_matb}} & 
\begin{tabular}{l c}  
\footnotesize{$\overleftarrow{\mathbf{W}}_\mathbf{x} = \mathbf{A}^H\,\overleftarrow{\mathbf{W}}_\mathbf{y} \, \mathbf{A}, \; \overleftarrow{\mathbf{W}}_\mathbf{x} \, \overleftarrow{\mathbf{m}}_\mathbf{x} = \mathbf{A}^H \, \overleftarrow{\mathbf{W}}_\mathbf{y} \, \overleftarrow{\mathbf{m}}_\mathbf{y}$} & \footnotesize{\numberTblEq{matb_v}} 
\end{tabular}
\\ [-0.1ex] \hline \\ [-2.4ex]

\raisebox{-0.5\height}{\includegraphics[scale=0.5]{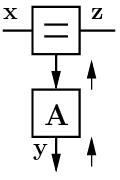}} &  
\begin{tabular}{l c}
\footnotesize{$\overrightarrow{\mathbf{V}}_\mathbf{z} = \overrightarrow{\mathbf{V}}_\mathbf{x} -  \overrightarrow{\mathbf{V}}_\mathbf{x} \mathbf{A}^H \mathbf{B} \mathbf{A} \overrightarrow{\mathbf{V}}_\mathbf{x} \;\quad\quad\quad$} & \footnotesize{\numberTblEq{composite_forw_v}} \\

\footnotesize{$\overrightarrow{\mathbf{m}}_\mathbf{z} = \overrightarrow{\mathbf{m}}_\mathbf{x}+\overrightarrow{\mathbf{V}}_\mathbf{x} \mathbf{A}^H \mathbf{B} (\mathbf{m}^\uparrow_\mathbf{y} - \mathbf{A}  \overrightarrow{\mathbf{m}}_\mathbf{x}) \;\quad\quad\quad$} & \footnotesize{\numberTblEq{composite_forw_m}} 
\\ 
\footnotesize{$\mathbf{B} = (\mathbf{V}^\uparrow_\mathbf{y} + \mathbf{A} \overrightarrow{\mathbf{V}}_\mathbf{x} \mathbf{A}^H)^{-1} \;\quad\quad\quad$} & \footnotesize{\numberTblEq{composite_defnB}} \\ 
\end{tabular}
 \\ [-0.5ex] \hline \\ [-2.1ex]
 
\raisebox{-0.4\height}{\includegraphics[scale=0.5]{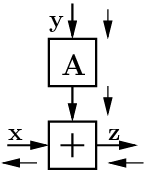}} &  
\begin{tabular}{l c}
\footnotesize{$\overleftarrow{\mathbf{W}}_\mathbf{x} = \overleftarrow{\mathbf{W}}_\mathbf{z} -  \overleftarrow{\mathbf{W}}_\mathbf{z} \mathbf{A} \mathbf{C} \mathbf{A}^H \overleftarrow{\mathbf{W}}_\mathbf{z} \quad\quad\quad$} & \footnotesize{\numberTblEq{composite_back_v}} \\

\footnotesize{$\overleftarrow{\mathbf{W}}_\mathbf{x} \overleftarrow{\mathbf{m}}_\mathbf{x} = (\mathbf{I}- \overleftarrow{\mathbf{W}}_\mathbf{z} \mathbf{A} \mathbf{C}  \mathbf{A}^H) * $} & \\  \footnotesize{$\quad\quad\quad\quad\quad\quad (\overleftarrow{\mathbf{W}}_\mathbf{z} \overleftarrow{\mathbf{m}}_\mathbf{z} -\overleftarrow{\mathbf{W}}_\mathbf{z} \mathbf{A} \mathbf{m}^\downarrow_\mathbf{y}) \quad\quad\quad$}  & \footnotesize{\numberTblEq{composite_back_m}} 
\\ 
\footnotesize{$\mathbf{C} = (\mathbf{W}^\downarrow_\mathbf{y} + \mathbf{A}^H \overleftarrow{\mathbf{W}}_\mathbf{z} \mathbf{A})^{-1} \;\;\:\quad\quad\quad\quad\quad\quad\quad $} & \footnotesize{\numberTblEq{composite_defnC}} 
\end{tabular}
\\ [-0.2ex] \hline  
\end{tabular}
\end{table}

\bibliography{IEEEabrv,MIMO_LMMSE}

\end{document}